\documentclass[prd,aps,floatfix,superscriptaddress,twocolumn]{revtex4-1}

\usepackage{amssymb}
\usepackage{mathtools} 
\usepackage[pdftex]{graphicx}
\usepackage[usenames, dvipsnames, svgnames, table]{xcolor}
\usepackage[colorlinks=true, citecolor=Blue,
            linkcolor=BrickRed, urlcolor=ForestGreen]{hyperref}
\usepackage{txfonts}
\usepackage{mathrsfs}
\usepackage{bm}
\usepackage{multirow}

\begin{document}   

\title{Towards the design of gravitational-wave detectors for probing neutron-star physics} 

\author{Haixing Miao} 
\affiliation{School of Physics and Astronomy, and Institute of Gravitational Wave Astronomy, 
University of Birmingham, Edgbaston, Birmingham B15 2TT, United Kingdom}
\author{ Huan Yang }
\email{hyang@perimeterinstitute.ca}
\affiliation{Perimeter Institute for Theoretical Physics, Waterloo, ON N2L2Y5, Canada}
\affiliation{University of Guelph, Guelph, ON N2L3G1, Canada}
\author{Denis Martynov}
\affiliation{LIGO, Massachusetts Institute of Technology, Cambridge, MA 02139, USA}
\affiliation{School of Physics and Astronomy, and Institute of Gravitational Wave Astronomy,   
University of Birmingham, Edgbaston, Birmingham B15 2TT, United Kingdom}

\begin{abstract}
The gravitational waveform of merging binary neutron stars encodes information about  extreme states of matter. 
Probing these gravitational emissions requires the gravitational-wave 
detectors to have high 
sensitivity above 1 kHz. Fortunately for current advanced detectors, there is a sizeable gap between the quantum-limited sensitivity and the classical noise at high 
frequencies. Here we propose a detector design that closes such a
gap by reducing the high-frequency quantum noise with 
an active optomechanical filter, frequency-dependent squeezing, and high optical power. 
The resulting noise level from 1 kHz to 4 kHz approaches the 
current facility limit and is a 
factor of 20 to 30 below the design of existing advanced detectors. 
This will allow for precision measurements of (i) the post-merger signal of 
the binary neutron star, (ii) late-time inspiral, merger, and ringdown of low-mass 
black hole-neutron star systems, and possible detection of (iii) high-frequency modes during supernovae explosions. This design tries to maximize the science return of current facilities by achieving a sensitive frequency band that is complementary to the longer-baseline third-generation detectors: the10 km Einstein Telescope, and 
40 km Cosmic Explorer. We have highlighted the main technical challenges towards 
realizing the design, which requires dedicated research programs. If demonstrated in 
current facilities, the techniques can be transferred to new facilities with longer baselines. 
\end{abstract}

\maketitle 

\section{Introduction}

The discovery of a binary neutron star merger event in August 2017 has marked the birth 
of multi-messenger astronomy including gravitational waves (GW)\,\cite{GW170817}. Soon, we also expect detections of black hole-neutron star mergers. These coalescence 
events, with matter involved, produce copious electromagnetic (EM) radiation in addition to GW emission, 
e.g., short-gamma-ray burst and kilonovas\,\cite{li1998transient,rosswog2005mergers,metzger2010electromagnetic}. The joint observation of both GW and EM signals in GW170817 
has already led to fundamental breakthroughs in understanding neutron star
the equation of state, 
the origin of short-gamma-ray bursts and heavy elements in our universe\,\cite{abbott2017multi,
GW170817GRB, abbott2017estimating}. 

Current advanced GW detectors are primarily sensitive to the low-frequency, inspiral stage of a binary neutron star
(BNS) merger, unless the source distance 
is within of the order of 10 Mpc. The merger processes, however, contain
rich information on the physics of nuclear matters under extreme conditions. In order to 
probe them through GW observations, we need better 
detector sensitivity above 1 kHz to resolve various spectral signatures, 
such as the main peak\,\cite{yang2017gravitational}, 
sub-dominant mode features\,\cite{Clark:2014wua}, and 
the one-arm instability\,\cite{Paschalidis:2015mla,East:2015vix,
Lehner:2016wjg}. Highly eccentric BNS inspirals may also emit GWs at f-mode frequency above 1 kHz \cite{Yang:2018bzx}. Given a 1.35 $M_{\odot}$-1.35 $M_{\odot}$ BNS
merger at a distance of $\sim50\,\rm Mpc$, the typical magnitude 
of the post-merger strain from 
$1\,\rm kHz$ to 4\,kHz is of the order of $2.0 \times 10^{-22}$. 
If the detector can achieve a sensitivity of $5.0 \times 10^{-25} \,{\rm Hz}^{-\frac12}$, which defines our target, the signal-to-noise ratio (SNR) will be around 20 and 
precise measurement of the 
BNS post-merger waveform will become possible. Not only can we
 explore the merger physics, but we may also make an independent determination of the Hubble constant\,\cite{Messenger:2013fya, schutz1986determining,PhysRevD.85.023535,messenger2012measuring}. 
Furthermore, such a level of sensitivity allows us to detect 
signals from merging NS and low-mass black hole (BH) binaries, 
to perform spectroscopy of high-frequency modes during supernovae explosions and magnetar giant flares,
and possibly to detect the stochastic GW background from BNS merger remnants. 

To achieve the target sensitivity, we need to reduce detector noise at high frequencies. 
The dominant noise comes from the counting statistics of photons, also known as the shot noise. Its typical 
magnitude, in terms of noise spectral density, is approximately equal to 
\begin{equation}\label{eq:shotnoise}
10^{-24} \,{\rm Hz}^{-\frac12}\left( \frac{1\,{\rm MW}}{P_{\rm arm}}\right)^{\frac12} \left( \frac{\lambda}{1064\,{\rm nm}}\right)^{\frac12}
\left(\frac{\Delta f}{3\, {\rm kHz}}\right)^{\frac12} \left(\frac{10}{e^{2r}}\right)^{\frac12}\,, 
\end{equation}
where $P_{\rm arm}$ is the power inside the interferometer arm cavity, $\lambda$ is the laser wavelength, 
$\Delta f$ is the detection bandwidth, and $e^{2r}=10$ for 10dB squeezing, if the squeezed light is used. 
There are three approaches to reduce the shot noise\,\cite{Rana:Review2013}: 
 increasing the power, injecting squeezing\,\cite{geo_squeezed, lisa_squeezed, Schnabel2017}\,, 
and signal recycling\,\cite{Meers1988, Chen1, future_detune_src, Thuring2007}. 
The third approach, however, is constrained by a tradeoff between the bandwidth and the peak sensitivity---higher peak sensitivity requires 
narrowing the detection bandwidth, as we can see from Eq.\,\eqref{eq:shotnoise}. 
We consider using an active optomechanical filter to surpass this constraint\,\cite{Miao2015a,Page2017}. Together with high optical power and 
frequency-dependent squeezing\,\cite{Kimble02, Oelker2016}, this enables us to
reach around $5.0 \times 10^{-25} \,{\rm Hz}^{-\frac12}$ from 1\,kHz to 4\,kHz.

\section{Detector design}
 
\begin{figure*}[tb]
\includegraphics[width=0.53\linewidth]{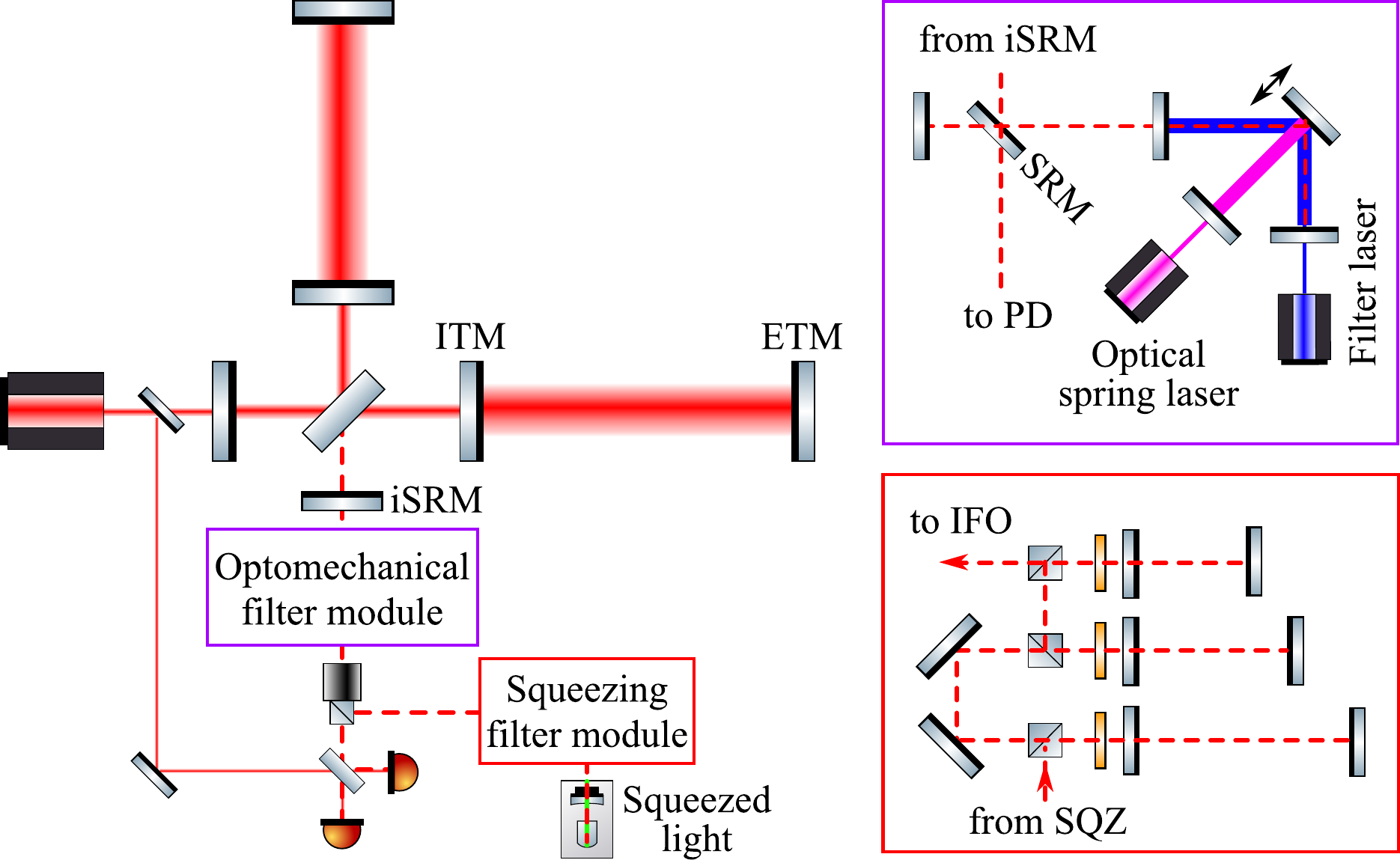}
\quad\quad
\includegraphics[width=0.43\linewidth]{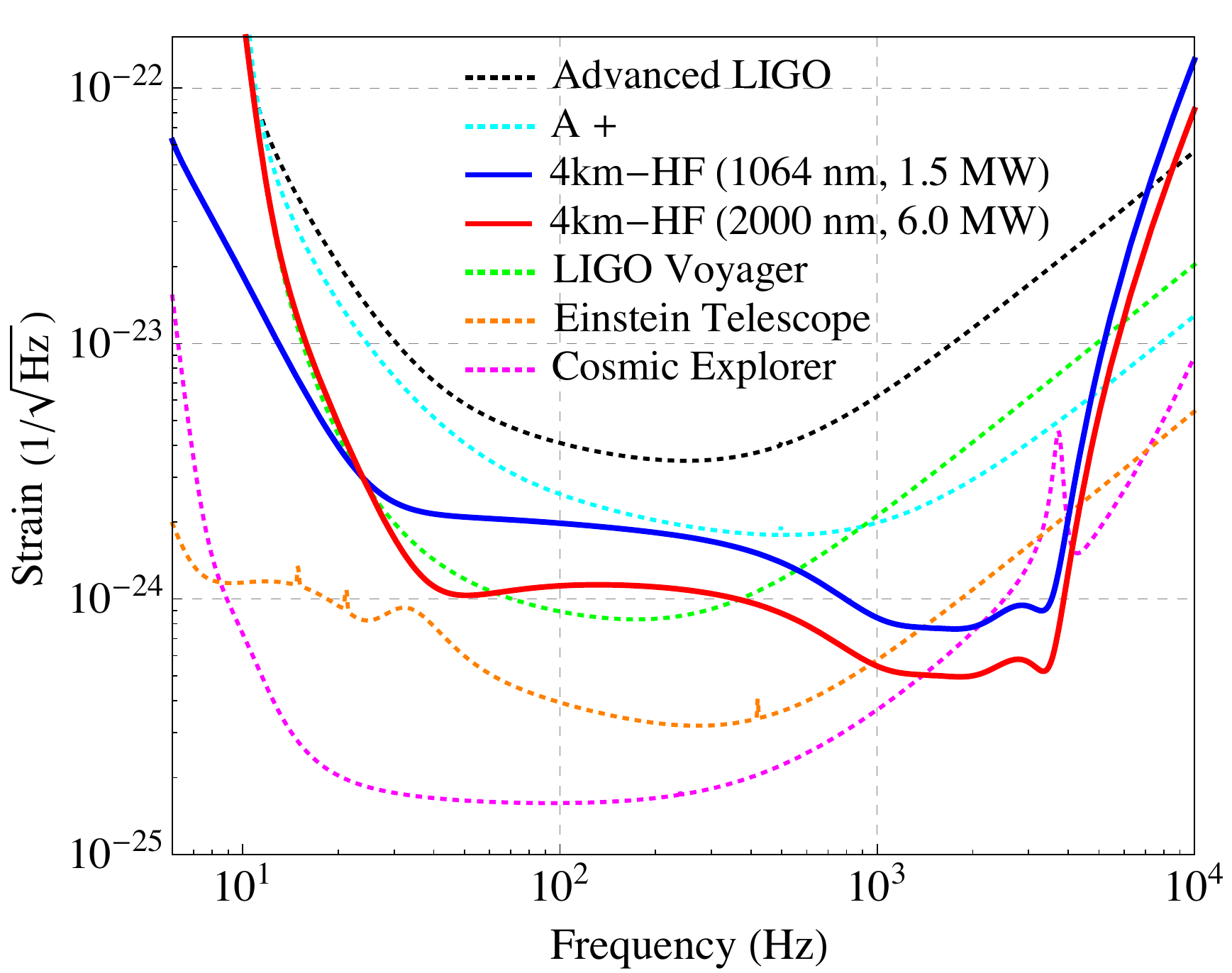}
\caption{Schematics showing the detector design  (left) and the resulting 
sensitivity (right). The blue curve shows a possible
intermediate step. The target 
sensitivity curve in red is around a factor of 30 below 
Advanced LIGO design at 3 kHz: a factor of 3 from 10\,dB squeezing, 
2 from high power, and 5 from detuned signal recycling with the 
optomechanical filter extending the improvement around the detune frequency. 
Sensitivity curves of Advanced LIGO plus (A+)\,\cite{Miller2015}, LIGO Voyager\,\cite{Adhikari2017}, 
 Einstein Telescope\,\cite{Punturo10a},   
and Cosmic Explorer\,\cite{abbott2017exploring} are shown as references.
A list of acronyms: PD for photodiode, IFO for interferometer, 
SQZ for squeezed light, and HF for high frequency detector. }
\label{fig:config}
\end{figure*} 

\begin{table}[!h]
\begin{ruledtabular}
\begin{tabular}{c|lc}
 &  Parameters & Values \\
 \hline
 \parbox[t]{3mm}{\multirow{9}{*}{\rotatebox[origin=c]{90}{Interferometer}}} 
 & arm length $L_{\rm arm}$ & 4 km  \\
 & arm cavity power $P_{\rm arm}$   & 6.0 MW (1.5MW) \\
 & test mass $M$ & 200 kg \\
 & laser wavelength $\lambda$ & 2000 nm (1064 nm) \\
 & temperature & 120 K (295 K) \\
 & SRM transmission & 3750 ppm\\
 & signal-recycling detune $\Delta_{\rm SR}$ & 1.5 kHz \\
 & SRC round-trip loss $\epsilon_{\rm SRC}$ & $\leq$ 150 ppm (300 ppm) \\
 & output loss & $\leq$ 3\% (5\%) \\
 \hline
\parbox[t]{3mm}{\multirow{17}{*}{\rotatebox[origin=c]{90}{Optomechanical filter}}} 
 & oscillator mirror mass $m$ &  5 mg \\ 
 & mirror radius, thickness & 1.4 mm, 0.35 mm \\ 
 & loss angle of substrate, coating  & $1.0 \times 10^{-9}$, $2.0 \times 10^{-6}$  \\
 & suspension quality factor & $3.0 \times 10^6$ \\
 & optical spring (OS) frequency $\omega_{\rm OS}$ & 12 kHz  \\
 & cavity length & 4.3 m \\
 & cavity bandwidth & 1.4 kHz \\
 & beam radius & 0.52 mm \\
 & resonating power & 338 W (180 W) \\
 & round-trip loss & $\leq$ 5 ppm (10 ppm) \\
 & laser wavelength for OS cavity $\lambda_{\rm OS}$ & 1064 nm \\
 & OS photodiode efficiency $\eta_{\rm OS}$ & $\ge$ 0.999 \\
 & OS cavity length & 10 cm \\
 & OS cavity bandwidth $\gamma_{\rm OS}$, detune $\Delta_{\rm OS}$ & 60 kHz, 0.9 MHz\\
 & OS cavity resonating power $P_{\rm OS}$ & 680 W \\
 & OS cavity round-trip loss $\epsilon_{\rm OS}$ & $\leq$ 1 ppm \\
 & temperature $T_{\rm env}$ & 16 K \\
 \hline 
\parbox[t]{3mm}{\multirow{6}{*}{\rotatebox[origin=c]{90}{SQZ filter}}} & squeezing (observed) & 10 dB \\ 
  & filter cavity 1 (bandwidth, detune) & 4.66 Hz, $-42.6$ Hz \\
 & filter cavity 2  & 197 Hz, 3409 Hz  \\ 
 & filter cavity 3  & 355 Hz, 1107 Hz \\
 & filter cavity 4  & 510 Hz, $-1920$ Hz 
\end{tabular}
\end{ruledtabular}
\caption{Parameters of the design.  Values in the parentheses are those used 
in producing the blue curve in Fig.\,\ref{fig:config}.}
\label{tab:parameters}
\end{table}

The proposed configuration is similar to that of Advanced LIGO\,\cite{design_aligo}, Advanced VIRGO\,\cite{aVIRGO} and KAGRA\,\cite{Aso13}, which consists of power recycling, signal recycling,
and arm cavities. We consider two cases for the interferometer: the first one, 
as an intermediate step, assumes a 1064\,nm laser and LIGO-LF classical noise level\,\cite{Yu2017}, and 
the second one, which achieves the target sensitivity, employs 2000\,nm 
and has the classical noise budget of LIGO Voyager\,\cite{Adhikari2017}. 
Envisioning progress in the capability of handling high power, 
we assume that the arm cavity powers for both cases are doubled compared to 
their original design, i.e., 1.5 MW and 6.0 MW, respectively. The radiation pressure 
effect of 6 MW at 2000 nm is equivalent 
to 3.0 MW at 1064 nm. 

Compared to current detectors, the difference in the design comes 
from the configuration of the signal recycling cavity (SRC) 
as shown in Fig.\,\ref{fig:config}. 
We introduce an internal signal recycling mirror (iSRM) to form
an impedance matched cavity with the input test mass (ITM) mirror. The advantage is
that the GW signal is not affected by the narrow bandwidth of the arm cavity.
However, optical loss in the central beam splitter and also the ITM substrate
are resonantly enhanced, which puts a hard bound on the 
sensitivity\,\cite{Miao2017}. For example,
100 ppm loss ($\epsilon_{\rm SRC}=10^{-4}$) inside 
SRC will lead to a sensitivity limit around $10^{-25}\, {\rm Hz}^{-\frac12}$ at 2\,kHz:\begin{equation}
10^{-25}\,{\rm Hz}^{-\frac12}\left( \frac{6\,{\rm MW}}{P_{\rm arm}}\right)^{\frac12} \left( \frac{\lambda}{2000\,{\rm nm}}\right)^{\frac12}\left( \frac{\epsilon_{\rm SRC}}{10^{-4}}\right)^{\frac12} \left( \frac{0.028}{T_{\rm ITM}}\right)^{\frac12}\,.
\end{equation} 
To reach the target sensitivity, we require the SRC loss to be less than 150\,ppm for 2000\,nm wavelength and 300\,ppm for 1064\,nm, while also setting
 the ITM transmission $T_{\rm ITM}$ to be 0.028, which is two times larger than Advanced LIGO.

In addition to the iSRM, 
an optomechanical filter module\,\cite{Miao2015a} is added to the SRC. 
This module compensates for the phase lag acquired by signal sidebands
when propagating in the interferometer arm, and results in a broadband 
resonance of the signal. It consists of a cavity-assisted 
optomechanical device\,\cite{Chen2013, Aspelmeyer2014}---an optical cavity with a movable mirror as the mechanical oscillator 
(highlighted in Fig.\,\ref{fig:config} with a double-headed arrow) and 
an additional laser field as the pump (named the filter laser) which induces a radiation 
pressure coupling between the oscillator and the signal sideband field. 
The filter laser frequency is higher than the carrier frequency $\omega_0$ 
of the main interferometer by the sum of the mechanical oscillator frequency $\omega_m$ 
and the signal-recycling detuning frequency $\Delta_{\rm SR}$ of the interferometer. The 
filter is operating in the unstable regime with the mechanical oscillator having 
a negative damping rate of $\gamma_{\rm opt}$ due to the optomechanical 
interaction, and the entire system is stabilised 
with a feedback control. The resulting open-loop transfer function for the signal 
sideband field at $\omega_0+\Omega$ is approximately given by (the exact transfer function 
is used for the noise analysis that produces the noise curves in Fig.\,\ref{fig:config}): 
\begin{equation}
\frac{\gamma_{\rm opt}-i(\Omega-\Delta_{\rm SR})}{\gamma_{\rm opt}+i(\Omega-\Delta_{\rm SR})}
\approx e^{-2i(\Omega-\Delta_{\rm SR})/\gamma_{\rm opt}}\,, 
\end{equation}
where $\Omega=2\pi f$ is the 
GW signal frequency in $\rm rad/s$.
When $\gamma_{\rm opt}$ is tuned to be equal to $c/L_{\rm arm}$ by changing the 
filter laser power, we can achieve the desired negative dispersion to cancel the propagation 
phase of the sideband, which is equal to $2(\Omega-\Delta_{\rm SR}) L_{\rm arm}/c$ .

One critical issue of the optmechanical filter is the fluctuation of the mechanical 
oscillator around its resonance as it is directly down-converted to the GW frequency 
band. We propose to implement the mechanical oscillator using
a low-loss quasi-monolithic suspension~\cite{sus_aston} and 
a milligram-scale mirror to achieve a low suspension thermal noise. 
For the phase compensation to work in 
the kHz regime, the oscillator frequency needs to be larger than the filter cavity 
bandwidth which in turn shall be larger than the kHz GW signal frequency---the so-called 
resolved sideband regime, as discussed in Ref.\,\cite{Miao2015a}. 
We increase the oscillator frequency from its bare value $\omega_m$ 
to $\omega_m'=(\omega_m^2+\omega^2_{\rm OS})^{1/2} \approx \omega_{\rm OS}$ by creating an optical spring at $12$ kHz with an auxiliary optical 
cavity\,\cite{Corbitt:2007lr}. The optical spring cavity
 is shown as the additional linear cavity normal to 
the oscillator mirror in Fig.\,\ref{fig:config}. There are five primary sources of noise:
suspension thermal noise, quantum radiation pressure noise from the optical spring,
coating thermal noise, substrate Brownian noise, and thermoelastic noise. To achieve 
the desired sensitivity level of the order of $10^{-25}/\sqrt{\rm Hz}$, these noises need 
to be suppressed down to a low level, the detail of which is presented in 
Appendix\,\ref{app:A}. Here we summarize the main challenges as follows. The 
total optical loss of the optical spring cavity should be below 
1\% to cancel the radiation pressure noise\,\cite{Korth2013a}, which requires
a high quantum efficiency photodiode~\cite{schnabel_qe}, adaptive mode matching~\cite{Liu2013}, and a high-quality optical cavity with
round trip loss of 1\,ppm. 
Achieving low mechanical dissipation in optical coatings and milligram-size mirror substrate requires a low-loss silicon mirror with 
crystalline coatings~\cite{Cole:2013} cooled down to 16\,K\,\cite{Harry2012}.

The squeezing filter module consists of a 
cascade of optical cavities to create frequency-dependent squeezing for 
reducing the quantum noise over a broad frequency band\,\cite{Kimble02, Oelker2016, 
Schnabel2017}. This design requires four filter cavities to realise the 
optimal squeezing angle\,\cite{Harms2003}. One is to reduce the
low-frequency radiation pressure noise. Its cavity bandwidth is around 5\,Hz, which implies a long cavity length due to the optical loss issue. If focusing on the sensitivity above 100 Hz, we can remove it; alternatively, we
can use the arm cavity filter idea\,\cite{Ma2017}. The other three are to undo the squeezing angle rotation at the detune frequency because of 
frequency dependence introduced by the detuned interferometer and the optomechanical filter. Their bandwidths are relatively large, and we can realise them using cavities of the order of ten meters. 

The resulting sensitivity curves are shown in Fig.\,\ref{fig:config}, and 
the parameters are summarized in Table\,\ref{tab:parameters}. The 
Supplemental Material provides additional information about the detailed 
noise budget and design. In the discussion that follows, we will present the
science case of this detector design.

\section{Scientific return}

In this section, we discuss several key aspects of the scientific return of this high-frequency detector design with its target sensitivity. They include spectroscopy of 
the BNS post-merger waveform, observation of low-mass BH-NS merger/post-merger dynamics, detection of the high-frequency stochastic GW background, and measurement of the Hubble constant in  
the absence of EM counterparts. These scenarios are all related to physical processes 
that are challenging to probe with low-frequency GW observations. 
In addition to these, more scientific cases are open for exploration in the high-frequency band from 1 kHz to 4 kHz, e.g., detecting high-frequency modes during
core-collapse supernovae explosions. 

\subsection{Precise spectroscopy of BNS post-merger dynamics}

Compared to their progenitors, BNS merger remnants contain hot, dense nuclear materials in a non-equilibrium state. Magneto-rotational instabilities and magnetic field winding likely boost the magnetic field to the order of $10^{15} \,{\rm G}$ level, with the remnant's rotation gradually slowing down. A precise GW spectroscopy of NS Helioseismology and the joint observation of EM radiations from, e.g., gamma-ray bursts and kilonovas, shall help us explore these most violent, matter-involving processes in the universe. 

Recently, progress has been made towards understanding the post-merger dynamics and the GW signature of the merger remnants\,\cite{baiotti2017binary}. The signal spectrum generically contains several characteristic peaks, which are related to the evolution of oscillation modes. These peaks sensitively depend on the equation of state (EOS), mass, initial rotation, and possibly magneto-hydrodynamic instability. Current simulations have not incorporated all physically relevant mechanisms, e.g., the neutrino radiation transport, which may be computationally prohibitive. Therefore the resulting post-merger waveforms still contain large theoretical uncertainties. 

\begin{figure}[tb]
\includegraphics[width=\columnwidth]{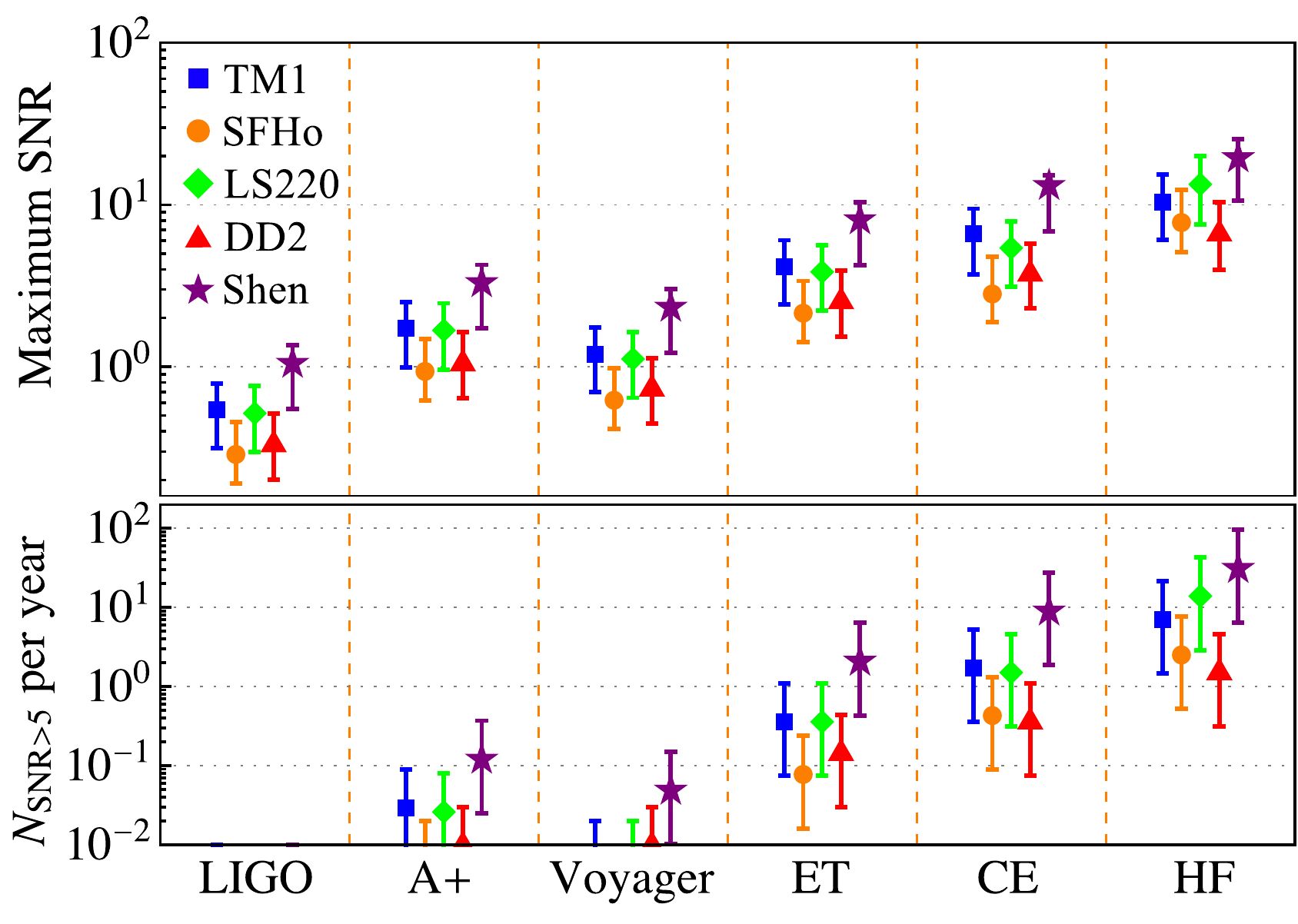}
\caption{The top panel shows the expected maximum SNR for detecting (2, 2) mode given different EOSs and detector sensitivities; the bottom panel shows the expected number of events with ${\rm SNR}\ge5$ with a one-year observation. The error bar corresponds to the $90\%$ confidence interval of the merger rate presented in Ref.\,\cite{GW170817}, with a most probable rate of ${\rm 1540 \,Gpc^{-3} yr^{-1}}$. }
\label{fig:pnsplot}
\end{figure}

Precise spectroscopy requires resolving individual modes. Here we focus on the dominant peak (2, 2) mode with indices $\ell=2$ and $m=2$ (spherical harmonics), which is one spectrum feature robustly determined in different simulations. Its waveform can be approximately modelled as a decaying sinusoid\,\cite{yang2017gravitational,bauswein2015exploring}. 
To quantify the detectability of the (2, 2) mode, we apply the Monte-Carlo (MC) method to generate mock data of BNS merger events and calculate the SNR for detecting it. 
The SNR is defined as
\begin{align}
{\rm SNR} = 2 \sqrt{\int^{f_{\rm high}}_{f_{\rm low}} df \frac{|h(f)|^2}{S_{hh}(f)}}
\end{align}
where $h(f)$ is the frequency-domain waveform (Fig.~\ref{fig:pnsplot} uses the (2,2) mode waveform and Fig.~\ref{fig:totalsnrplot} uses the total post-merger waveform) and $S_{hh}$ is the single-side noise spectral
density of the detector, which includes both the classical and quantum noise. 
The lower cut-off frequency $f_{\rm low}$ is set to be $1$ kHz and
 the upper cut-off frequency $f_{\rm high}$ is set to be $5$ kHz, 
 which is consistent with the numerical waveforms we have used.
We consider five representative EOSs: TM1, SFHo, LS220, DD2, and Shen, which are studied extensively in state-of-the-art simulations\,\cite{Palenzuela:2015dqa, kiuchi, Stergioulas:2011gd}, though other EOSs not discussed here may give rise to different post-merger waveforms. We assume the merger rate to be in the range ${\rm 320-4740 \,Gpc^{-3} yr^{-1}}$\,\cite{GW170817}.  In each MC realization, we randomly sample the source sky location, inclination angle, distance, and component mass which follows the distribution presented in Ref.\,\cite{ozel2016masses}. The event with the highest SNR is selected out at the end of each realization. 
We have computed $10^2$ MC realizations. In Fig.~\ref{fig:pnsplot}, we show the maximum SNR, averaged over all realizations, for given different detector sensitivities, 
and the expected number of events with ${\rm SNR} \ge 5$ in one year. We can see that the high-frequency detector can achieve high SNR with a good event rate. 

The same analysis can also be applied to other subdominant modes in the spectrum of the post-merger waveform \,\cite{Clark:2014wua,yang2017gravitational,bose2017neutron}. Measuring their frequencies and excitation amplitudes can provide crucial information about the NS EOS at different densities, the internal structure of the merger remnant, and the hydrodynamic processes. In addition, precise spectroscopy can also provide information about the redshift of the source, because the redshift dependences of the inspiral waveform and the post-merger waveform are different, which is in contrast to the binary black hole (BBH) waveform with the frequency being inversely proportional to the mass. For BNS, higher masses indicate smaller star sizes and a more compact remnant, which in turn leads to higher post-merger mode frequencies. Combined with the distance measurement based on the GW amplitude, this will allow determination of the Hubble constant, if we have an accurate understanding of NS EOS (based upon a few loud events with EM counterparts for calibration). As shown in the Supplemental Material, by detecting the (2, 2) mode alone, we could determine the Hubble constant to an accuracy of the order of $(0.1-0.4)/\sqrt{N}$ (where $N$ is the number of stacked events)  in the absence of EM counterparts. 

\begin{table}[b]
\centering
\begin{tabular}{c c c c c c}
\hline\hline
Type & $M_{\rm NS}$ ($M_\odot$) & $M_{\rm BH}/M_{\rm NS}$ & $\rm SNR_{\rm CE} $ & $\rm SNR_{\rm HF}$  & $f_{\rm cut} ($\rm kHz$)$\\
\hline
I & 1.35 &1.5 & 2.05 & 3.17 & 1.8 \\
I & 1.35 & 2 &  2.53 & 4.23 & 1.9 \\
II & 1.35 & 3 & 3.65 & 6.59 &  2.4  \\
III & 1.35 & 4 & 3.07 & 5.25 & 2.4  \\
III & 1.35 & 5 & 4.2 & 6.33 & 2.2 \\
\hline
\end{tabular}
\caption{SNRs for detecting different types of BH-NS mergers.}
\label{table:bhnssnr}
\end{table}

\subsection{ Neutron star-black hole binaries}

The discovery of GW150914 and the following observations of BBH mergers have drawn many researchers' interests to massive
 stellar-mass BHs because they emit stronger GWs. 
Low-mass BHs (LMBHs), however, are also interesting as they could come from entirely different progenitors\,\cite{yang2017can}. GW observations will be able to produce an independent measurement of the properties of such systems, including their mass/spin distributions and the cosmological evolution. More importantly, the coalescence between a NS and LMBH provides another exciting scenario for multi-messenger astronomy. It may not only generate GWs above 1\,kHz, but also emit EM radiations due to tidal disruption of the NS and mass accretion into the BH. Additionally, a massive BH may swallow its NS companion without producing energetic EM radiations. 

Based on the classification and simulations performed in Ref.\,\cite{shibata2009gravitational}, we compute the SNR for measuring (post-)merger waveforms of BH-NS binaries. The result is shown in Table \ref{table:bhnssnr} for both Cosmic Explorer (CE) and the high-frequency detector.  We assume a $\Gamma=2$ polytropic EOS, and a source distance of $50\,{\rm Mpc}$.  The cut-off frequency $f_{\rm cut}$ denotes the starting frequency of the (post-)merger waveform. The Type I, II, III waveforms correspond to three scenarios: the tidal disruption happening outside the innermost stable circular orbit (ISCO), inside ISCO, and no disruption, respectively.

\begin{figure}[!t]
\includegraphics[width=\columnwidth]{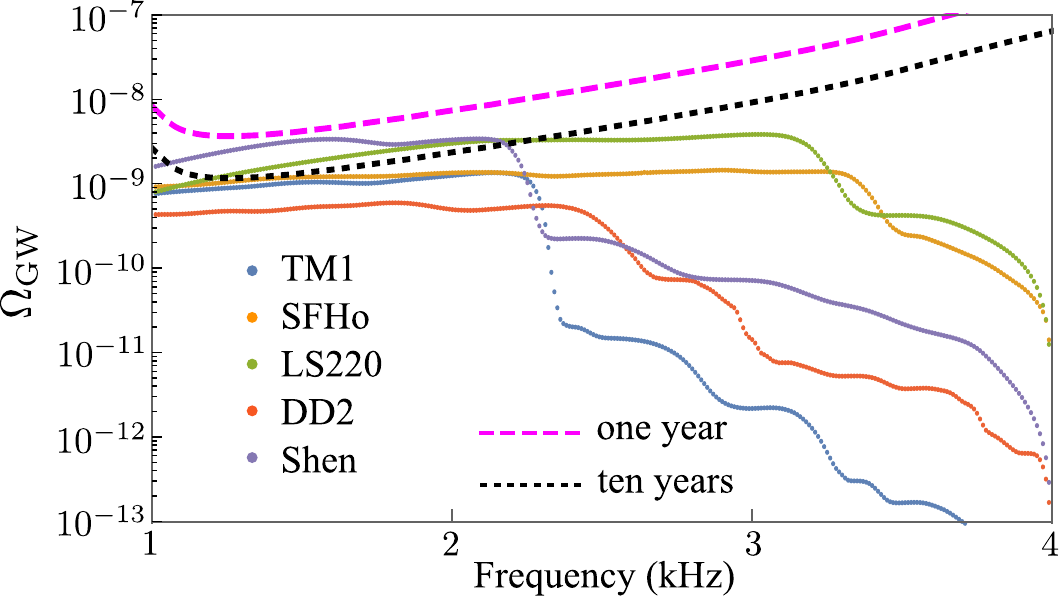}
\caption{Stochastic GW background produced by the post-merger GW emissions 
of BNSs with a merger rate of  ${\rm 1540 \,Gpc^{-3} yr^{-1}}$. Transition regions between different plateaus correspond to
dominant modes in the post-merger waveform, with the lowest-frequency 
one associated with the (2, 2) mode.  The magenta and black dashed lines are the ``power-law integrated"
sensitivity curves defined in Ref.\,\cite{thrane2013sensitivity}, with a one- and ten-year integration time, respectively.}
\label{fig:gwbplot}
\end{figure}

\subsection{Stochastic GW background}

At frequencies above $1 {\rm kHz}$, the component of the stochastic GW background (SGWB) contributed by binary BHs is significantly reduced\,\cite{GW170817}, and the SGWB is likely dominated by emissions from BNSs. The inspiral part of BNS waveforms cut off sharply below the BNS ISCO and the contact frequency is typically below $1.4\, {\rm kHz}$. Here we present a study on the contribution to the SGWB by BNS post-merger hydrodynamical oscillations\,\cite{Zhu:2012xw}. As shown 
in Fig.\,\ref{fig:gwbplot}, this part of the SGWB displays interesting plateau signatures associated with the main post-merger modes, which have not been discussed in the literature. It is the dominant source of the SGWB above the ISCO  frequency (cf.  Refs.\,\cite{cheng2017stochastic,marassi2011stochastic} to compare with estimations for magnetar emissions).
We find that the high-frequency detector, given a ten-year observation, can detect the SGWB for some EOS considered here with SNR  $\ge 2$. The cross-correlation search is not optimal for such non-Gaussian SGWB, and it may be possible to significantly improve the SNR with better search algorithms \cite{drasco2003detection,smith2017optimal}. We also would like to emphasize that 
this high-frequency window provides new opportunities to search for the SGWB of primordial or exotic origins, because of low astrophysical confusion noise.

\section{Conclusion and discussion}

This study illustrates both the challenges in improving the 
high-frequency sensitivity of GW detectors, and the exciting neutron-star 
science that is accessible by exploring this new frequency band. 
For the detector design, we have not exhausted all the possibilities for achieving the target sensitivity. The essential elements, 
e.g., squeezing, low optical loss, and high power, 
will, however, be shared by different designs. Additionally, 
using the optomechanical filter is not the only approach for broadening the
detection bandwidth, and there are others based upon atomic 
systems\,\cite{Wicht1997, Ma2015, Zhou2015}. 
Pushing the limit of these elements in different
approaches defines the direction of future research towards building high-frequency detectors. Once the techniques are well tested in current facilities, implementing them in facilities with longer arm lengths
will allow even better sensitivity and a much richer science return. 
For example, we may perform GW spectroscopy 
of the high-frequency part ($>1 \,{\rm kHz}$) of post-bounce supernovae oscillations up to several ${\rm Mpc}$.
Post-merger signals from BNSs at a larger distance could be observed,
which allows a more precise determination of the Hubble constant. We would also be able to probe high-frequency astrophysical processes and test predictions of General Relativity at cosmological distances, complementary to the information obtained from low-frequency GW observations. 

\section{Acknowledgments}
We would like to thank Chunnong Zhao, Alberto Vecchio, 
Luis Lehner, Yiqiu Ma, Andreas Freise, 
Conor Mow-Lowry, Rainer Weiss, Peter Fritschel, Matthew Evans, 
and members of the LSC AIC, MQM, and QN groups for 
fruitful discussions. We thank Andre Fletcher for proofreading 
a draft of this paper. H.M. is supported by UK STFC Ernest Rutherford 
Fellowship (Grant No. ST/M005844/11). 
H.Y. is supported by the Natural Sciences and Engineering Research Council of Canada, and in
part by Perimeter Institute for Theoretical Physics. Research at
Perimeter Institute is supported by the Government of Canada through
Industry Canada and by the Province of Ontario through the Ministry
of Research and Innovation. 
D.M. acknowledges the support of the NSF and the Kavli Foundation. 

\appendix

\section{Main interferometer noise budget}\label{app:A}

\begin{figure}[!h]
\includegraphics[width=\columnwidth]{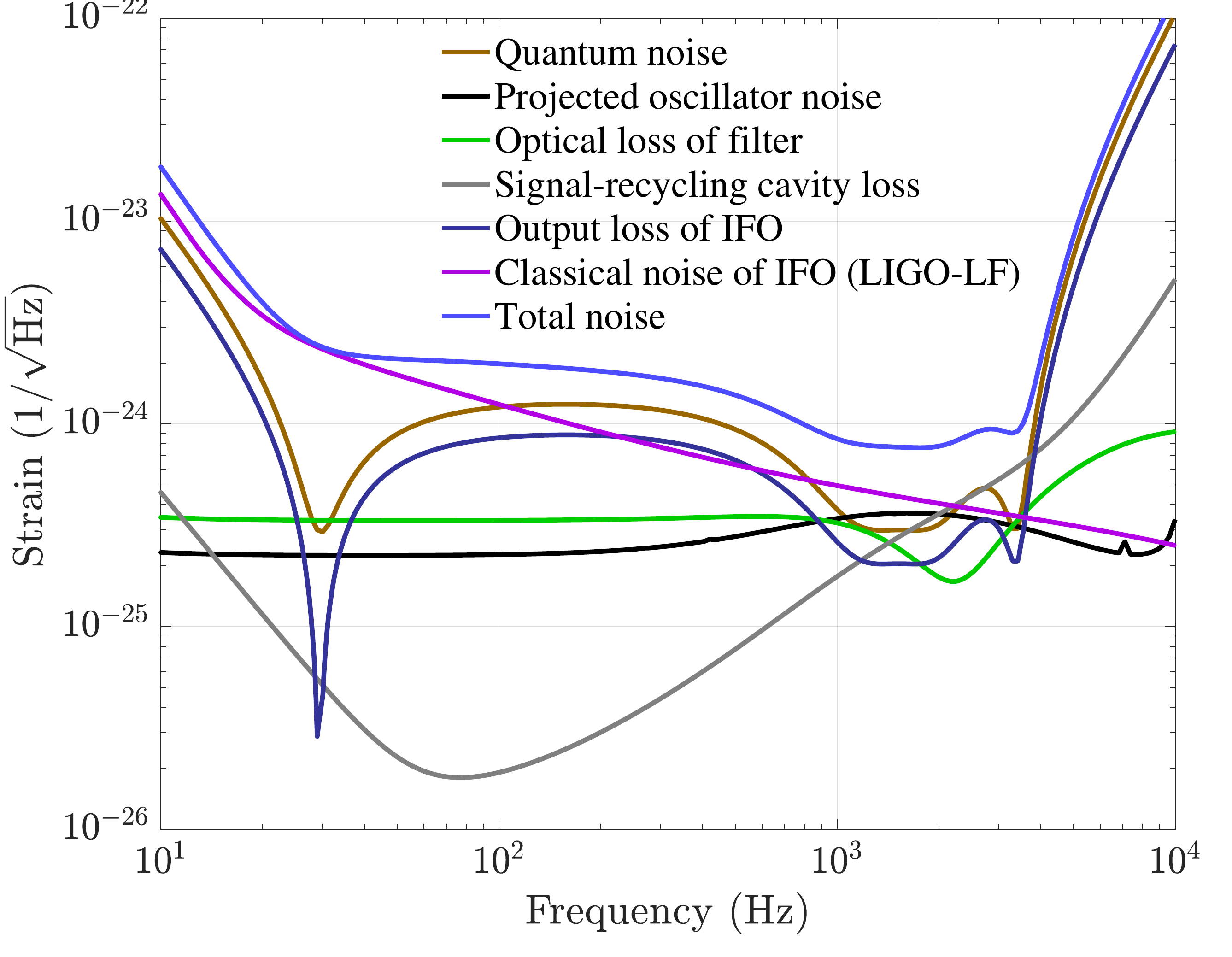}
\includegraphics[width=\columnwidth]{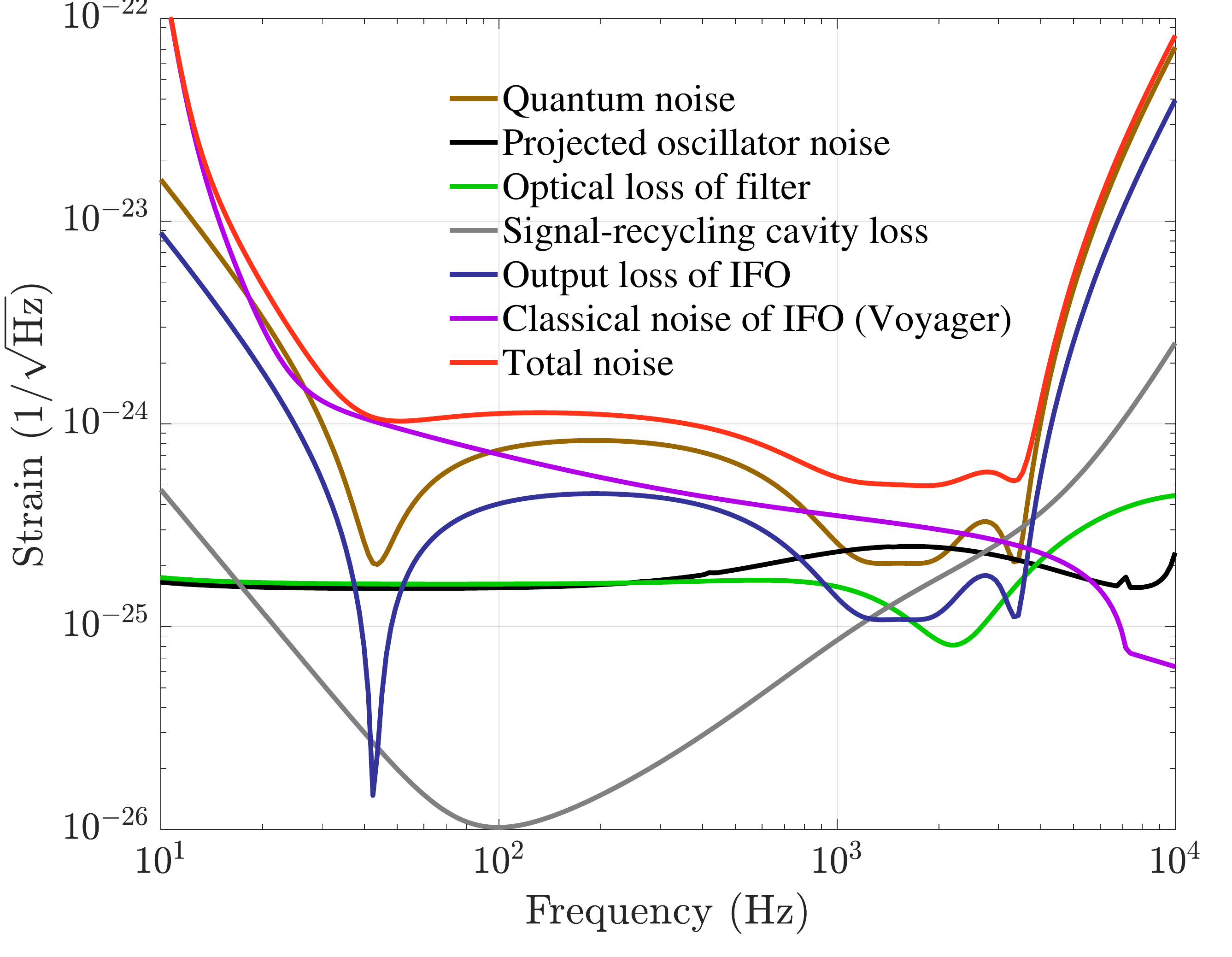}
\caption{The noise budget for the two cases of main interferometer: 
one assuming 1064 nm and LIGO-LF classical noise (upper panel), and the other 
assuming 2000 nm and  
LIGO Voyager classical noise (lower panel). The final sensitivity curves (total noise) were shown in Fig.\,1 of the main text. 
The projected oscillator noise comes from the mechanical
 oscillator in the optomechanical filter. We have also included the contributions from the optical loss in the filter cavity, the signal-recycling cavity, and the output.}
\label{fig:nb}
\end{figure}

\subsection{Noise budget}

The noise budget of our design is shown in Fig.\,\ref{fig:nb}. The classical noise from the main interferometer is based upon previous design studies: LIGO-LF and LIGO Voyager. We have also included noise contributions from the mechanical oscillator in the optomechanical filter, and from optical losses in the main components. We provide the details in the following two subsections.

\subsection{Realization of the optomechanical filter and its noise}

In this section, we discuss the experimental realization of the optomechanical filter and 
different noise sources associated with the mechanical oscillator. 

We propose to implement the oscillator using an $m = 5$\,mg mirror suspended with silicon fibers. The eigenfrequency is around $10$\,Hz with a typical quality factor of $10^6$ to $10^7$.
In order for the optomechanical filter to compensate the propagation phase of the signal sidebands properly, as shown 
in Ref.\,\cite{Miao2015a},  we need to be in the parameter regimes: $\omega_m \gg \gamma_f > \Omega$, where $\omega_m$
is the eigenfrequency of the oscillator, $\gamma_f$ is the filter cavity bandwidth, and  $\Omega$ is the sideband frequency (also the GW signal frequency). To meet this requirement, we shift the eigenfrequency up by using the optical spring effect in a detuned optical cavity, which is also called the optical dilution \cite{Corbitt:2007lr}.  The eigenfrequency changes to $\omega_m =\sqrt{\omega_{m0}^2+\omega_{\rm OS}^2} \approx \omega_{\rm OS}$ with 
$\omega_{m0}$ being the original frequency. The optical spring frequency $\omega_{\rm OS}$ is given by: 
\begin{equation}
\omega_{\rm OS}^2=\frac{32 \pi P_{\rm OS}}{\lambda_{\rm OS} T_{\rm OS}  m \,c}\frac{\kappa}{1+\kappa^2}\,. 
\end{equation}
Here $T_{\rm OS}$ is the transmission of the cavity input mirror, $P_{\rm OS}$ is the optical power resonating inside the cavity, 
and $\kappa = \Delta_{\rm OS} / \gamma_{\rm OS}$ is the ratio between the cavity
detuning and bandwidth. 
For the optimal performance of the optomechanical filter, we need to achieve $\omega_{\rm OS} /(2\pi) \geq 50$\,kHz to have phase compensation up to several kHz.  However, this requires significant optical power resonating in the cavity and extremely low optical loss. These requirements cannot be met using the current technology but might be achieved in the future. For the realization, 
we instead have $\omega_{\rm OS} /(2\pi) = 12$\,kHz. Such an optical spring frequency degrades the quantum-limited sensitivity of the GW detector in the frequency range $3-4$\,kHz compared with the ideal scenario. However, the resulting sensitivity is still significantly better than current detectors. 

The optical spring also allows us to dilute the suspension thermal noise around $\omega_{\rm OS}$. This is 
particularly important for the ultimate filter performance, since the mechanical motion 
at $\omega_{\rm OS}-\Delta_{\rm SR}
+ \Omega$ ($\Delta_{\rm SR}$ is the SR detuning frequency)  is directly down converted to  GW band at frequency $\Omega$.  However, the optical spring also amplifies the surface motion
of the oscillator from the coating and substrate thermal noises, as any displacement noise will appear as a force noise when multiplying the spring constant. 
In addition, there is extra quantum radiation pressure noise on the oscillator exerted by the same optical field that 
creates the optical spring. This noise can be reduced using the suppression technique, described in Ref.\,\cite{Korth2013a} based upon measurement feedback. There will still be some residual radiation pressure noise due to the optical loss of the cavity, 
the non-unity quantum efficiency of the photodetector, and frequency-dependent part of the noise that cannot be suppressed with feedback: 
\begin{equation}
\begin{split}
    & S_{FF}^{\rm rad} (\Omega) = \frac{2\hbar m \omega_{\rm OS}^2}{\kappa} 
    \left[\frac{\epsilon_{\rm OS}}{T_{\rm OS}} + (1-\eta_{\rm OS}) + \frac{\Omega^2}{\gamma_{\rm OS}^2} 
    \frac{\kappa^2}{(1+\kappa^2)^2}\right] \,,
\end{split}
\end{equation}
where $\epsilon_{\rm OS}$ is the round-trip optical loss of the cavity,  
and $\eta_{\rm OS}$ is the quantum efficiency of the 
photodiode.
In addition to the indirect path from the optical spring, coating thermal noise and substrate Brownian noise of the oscillator mirror also directly enters the signal beam in terms of displacement noises. The coherence of these two paths is ignored in our analysis, which is valid for the case when the laser wavelength for the optical spring cavity and the optomechanical filter are different. 

If we sum up all the noises above and view them as from a single dissipative process with viscous damping, we can use a single figure of merit to summarize the noise requirement: $T_{\rm env}/Q_m$. Here
$T_{\rm env}$ is the environmental temperature, 
and $Q_m$ is the equivalent quality factor around the oscillator eigenfrequency after accounting for all the noises. Since the oscillator noise degrades the detector sensitivity similar to the optical loss, we can convert this figure of merit into the magnitude of an effective optical loss, according to Eq.\,(13) in Ref.\,\cite{Miao2015a}:  
\begin{equation}
\epsilon_{\rm eff} = \frac{4 k_B}{\hbar \gamma_{\rm opt}}\left( \frac{T_{\rm env}}{Q_m}\right)\,, 
\end{equation} 
where $\gamma_{\rm opt}$ is set to be equal to $c/L_{\rm arm}$ for the phase compensation. With the 
optomechanical filter embedded inside the interferometer,
this loss can be viewed as an internal loss. 
According to Ref.\,\cite{Miao2017}, the ultimate sensitivity one can achieve with some internal optical loss 
of magnitude $\epsilon_{\rm int}$ is:
\begin{equation}
S_{hh}^{\epsilon}= \frac{\hbar\,  c^2 \epsilon_{\rm int}}{4 L_{\rm arm}^2 \omega_0
 P_{\rm arm}}\,,
\end{equation} 
where $P_{\rm arm}$ is the arm cavity power and $\omega_0=2\pi c/\lambda$ is the laser frequency.
To achieve 
a noise level of $S_{hh}^{1/2}=5.0\times 10^{-25}/\sqrt{\rm Hz}$ given 6 MW power and 2000 nm, the total internal
loss needs to be smaller than $10^4$ ppm, or equivalently,
\begin{equation}\label{eq:req}
\frac{T_{\rm env}}{Q_m} \le 1.4\times 10^{-9}\, \rm K\,.
\end{equation}

\begin{figure}[t]
\includegraphics[width=\columnwidth]{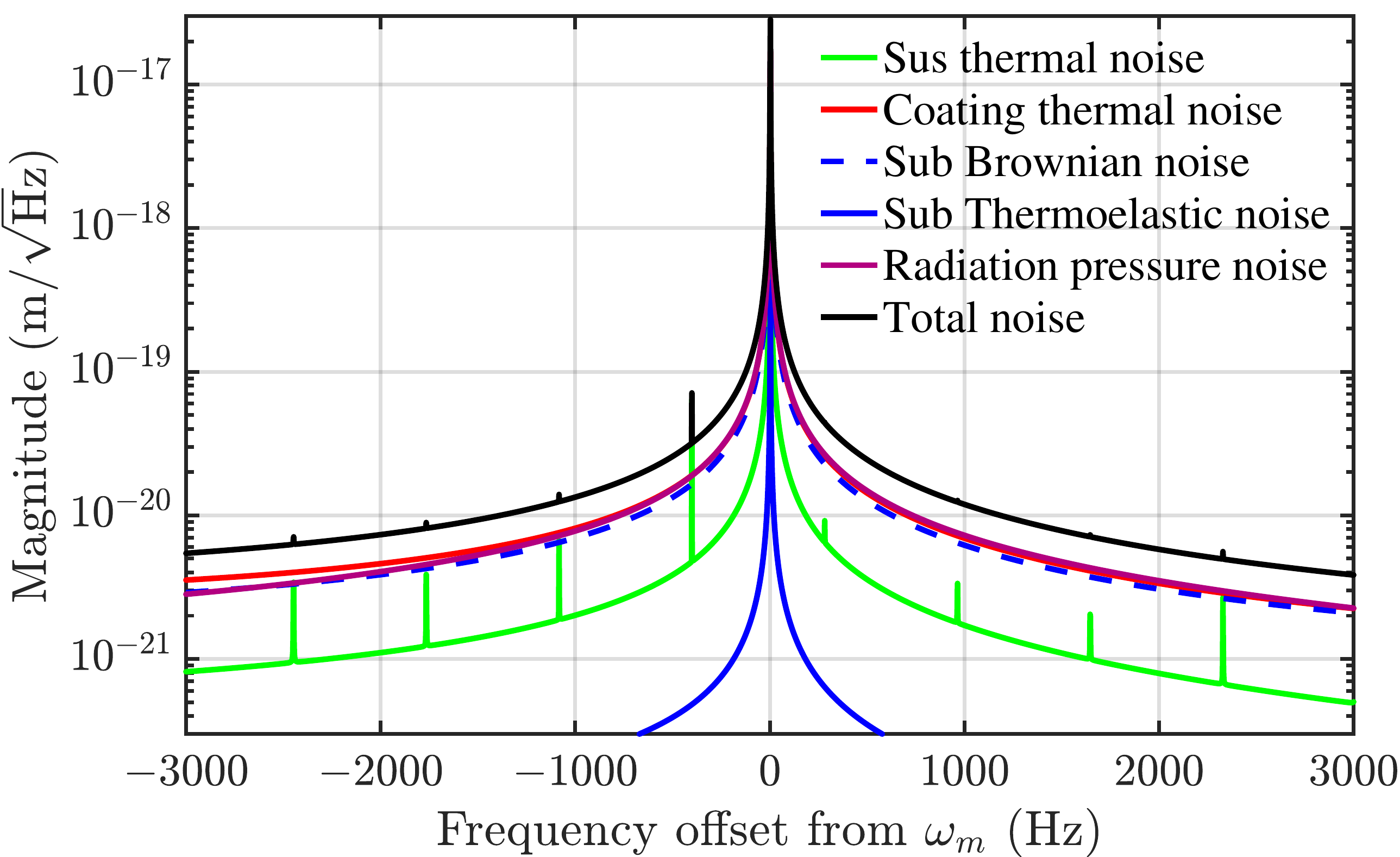}
\caption{Displacement noises of the mechanical oscillator near 12 kHz resonant frequency 
(the reference frequency) before the control laser for the optomechanical filter
 is turned on, which 
will modify the mechanical susceptibility. 
Configurations of the optical spring cavity and optomechanical
filter cavity were chosen to achieve a total noise level below the requirement in Eq.\,\eqref{eq:req}.}
\label{fig:os}
\end{figure}

Given the above considerations, we come up with the numbers in Table\,I of the main text and the spectral densities of different oscillator noises are shown in Fig.~\ref{fig:os}. 
We can satisfy the requirement Eq.\,\eqref{eq:req} in the frequency range 
$1 {\rm kHz}\leq \Omega / (2\pi)\leq 4$\,kHz given $\Delta_{\rm SR}/(2\pi)=1.5\,\rm kHz$. The equivalent $T_{\rm env}/Q_m$ is around $1.6\times 10^{-10}$ K. However, the effect of the internal loss is enhanced by a factor of 2; the quadrature that we measure for optimising the low-frequency sensitivity is not optimal for suppressing the effect of the internal loss. The total contribution from all noise sources in the oscillator to the final sensitivity is
approximately equal to $2.5\times 10^{-25}/\sqrt{\rm Hz}$ around 2 kHz in the case of 2000 nm, as shown in Fig.~\ref{fig:nb}.


\section{Quantum noise analysis}

\begin{figure}[b]
\includegraphics[width=\columnwidth]{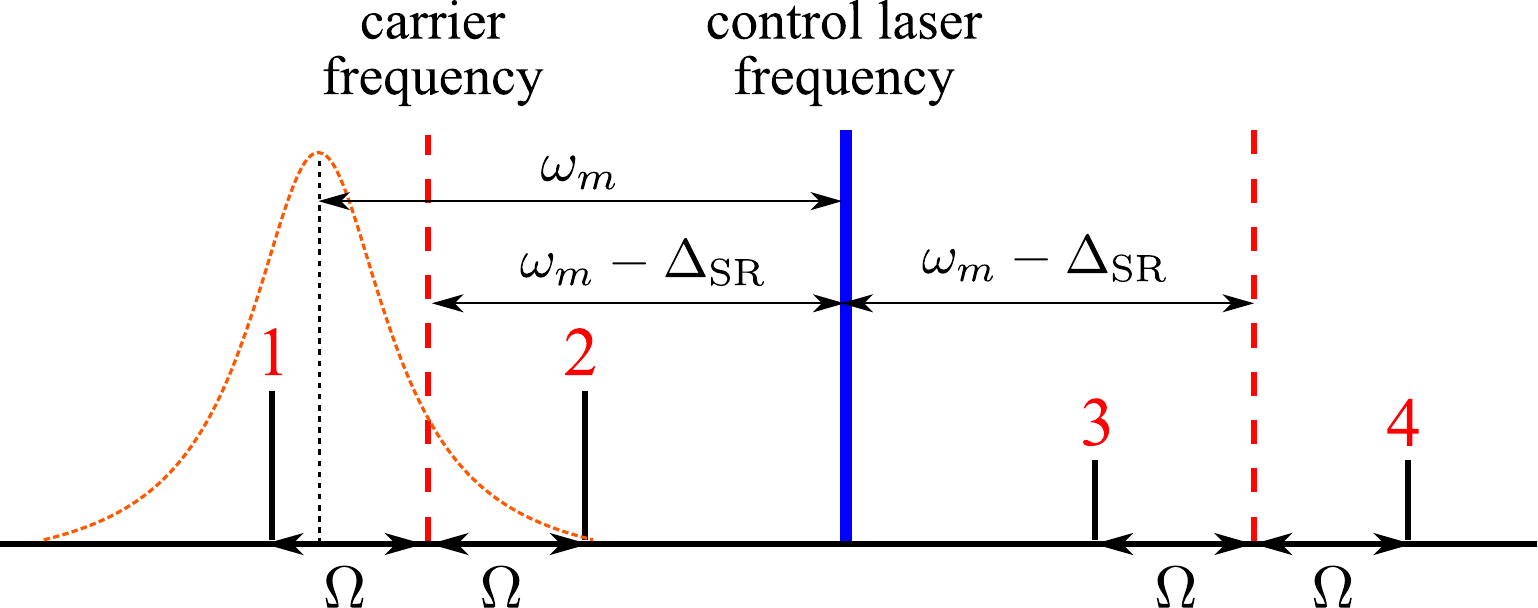}
\caption{A schematic showing the relevant four sidebands and their frequencies with 
respect to the carrier frequency $\omega_0$ and control laser frequency 
$\omega_0+\omega_m-\Delta_{\rm SR}$ ($\Delta_{\rm SR}$ is the signal
recycling detuning frequency of the interferometer). The control laser is detuned 
away from the optomechanical cavity resonant frequency by $\omega_m$; the 
resonance profile of the cavity is illustrated by the Lorentzian shape.}
\label{fig:sidebands}
\end{figure}

In this section, we will provide the details of how we perform the quantum 
noise analysis. The four relevant sideband fields are: (1) $\omega_0- \Omega$, 
(2) $\omega_0+ \Omega$,  (3) $\omega_0+ 2\omega_m-2\Delta_{\rm SR} - \Omega$ 
and (4) $\omega_0+ 2\omega_m-2\Delta_{\rm SR} + \Omega$, as illustrated 
in Fig.\,\ref{fig:sidebands}. The first two sidebands are our signal 
sidebands, which get mixed up due to the 
radiation pressure coupling between the 
light and test masses in the main interferometer; the latter two are idler sidebands, and
 come into the picture due to the radiation pressure coupling in 
the optomechanical filter cavity. 

\begin{figure}[t]
\includegraphics[width=0.7\columnwidth]{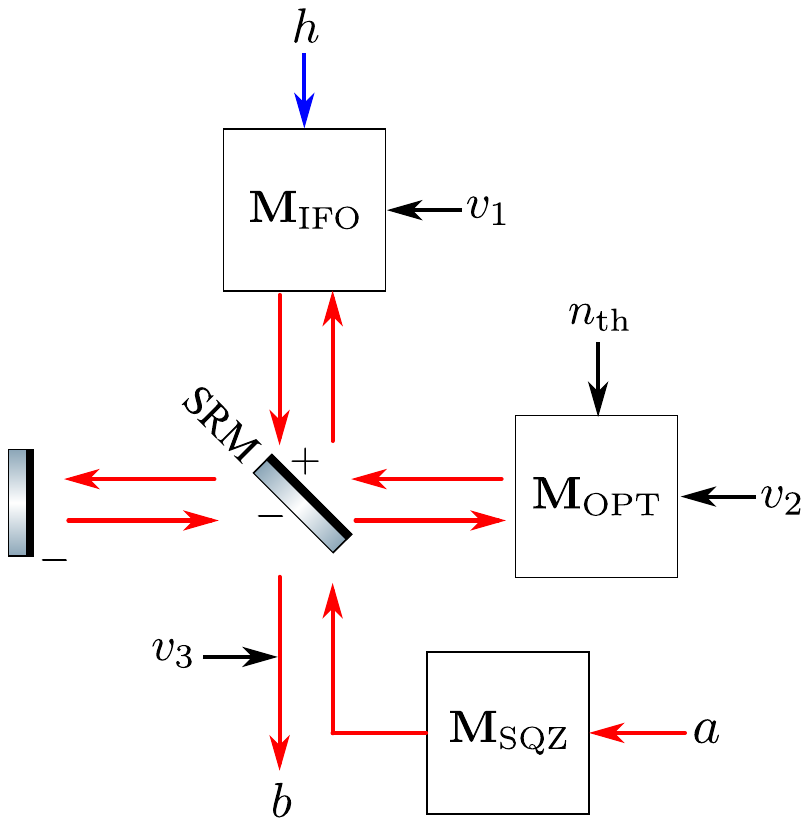}
\caption{A schematic showing the relevant fields, noises, GW signal, and 
the propagation 
transfer matrices involved in obtaining the input-output relation for 
calculating the sensitivity curve.}
\label{fig:io}
\end{figure}

The quantum-limited sensitivity can be obtained by using the standard formalism\,\cite{Kimble02,lrr-2012-5, Miao_2014}. It involves analyzing the propagation of these sidebands (or equivalent amplitude and phase quadratures) throughout the system, and obtaining the input-output relation at the differential port of the interferometer. 
The key transfer matrices involved in this analysis are illustrated in Fig.\,\ref{fig:io} 
(a simplified schematic of the configuration in Fig.\,1 of the main text): 
${\bf M}_{\rm IFO}$ is the matrix for the main interferometer, 
${\bf M}_{\rm OPT}$ for the optomechanical filter and ${\bf M}_{\rm SQZ}$  for the squeezing filter module and squeezed light source. Also in the same figure, 
$\bm a$ are the inputs of the sidebands and $\bm b$ are the outputs. 
To account for imperfections, we also include vacuum fields 
$\bm v_1$ (from optical loss in the 
IFO), $\bm v_2$ (loss in the
 optomechanical filter), $\bm v_3$ (loss at the output), and a thermal field
 $n_{\rm th}$ (thermal noise of the optomechanical filter). 
These additional fields also propagate throughout the system and finally add to the output $\bm b$. 
The mirror on the left-hand side of the SRM has a high reflectivity, which in the ideal 
case should be equal to 1, and it is to avoid introducing an open port that leads to
an additional vacuum noise input. 

Because we introduce an iSRM to form an impedance matched cavity with the 
ITM, the signal sideband part of ${\bf M}_{\rm IFO}$ is identical to that of a simple Michelson,  apart from the power being the arm cavity power and the mass being replaced by the reduced mass $M/2$ ($M=200\,$kg with our specifications). 
The idler sidebands around 
$\omega_0+2\omega_m-2\Delta_{\rm SR}$, 
when beating with the 
carrier at the frequency $\omega_0$, exert a radiation pressure on 
the test mass at frequency around $2\omega_m-2\Delta_{\rm SR}$ (tens of kHz).
This radiation pressure is at a much higher frequency than the test-mass pendulum frequency 
(around 1 Hz); we can, therefore, ignore their radiation pressure effect on the test mass, and view them as freely propagating through the arm cavity. 

\begin{figure}[b]
\includegraphics[width=\columnwidth]{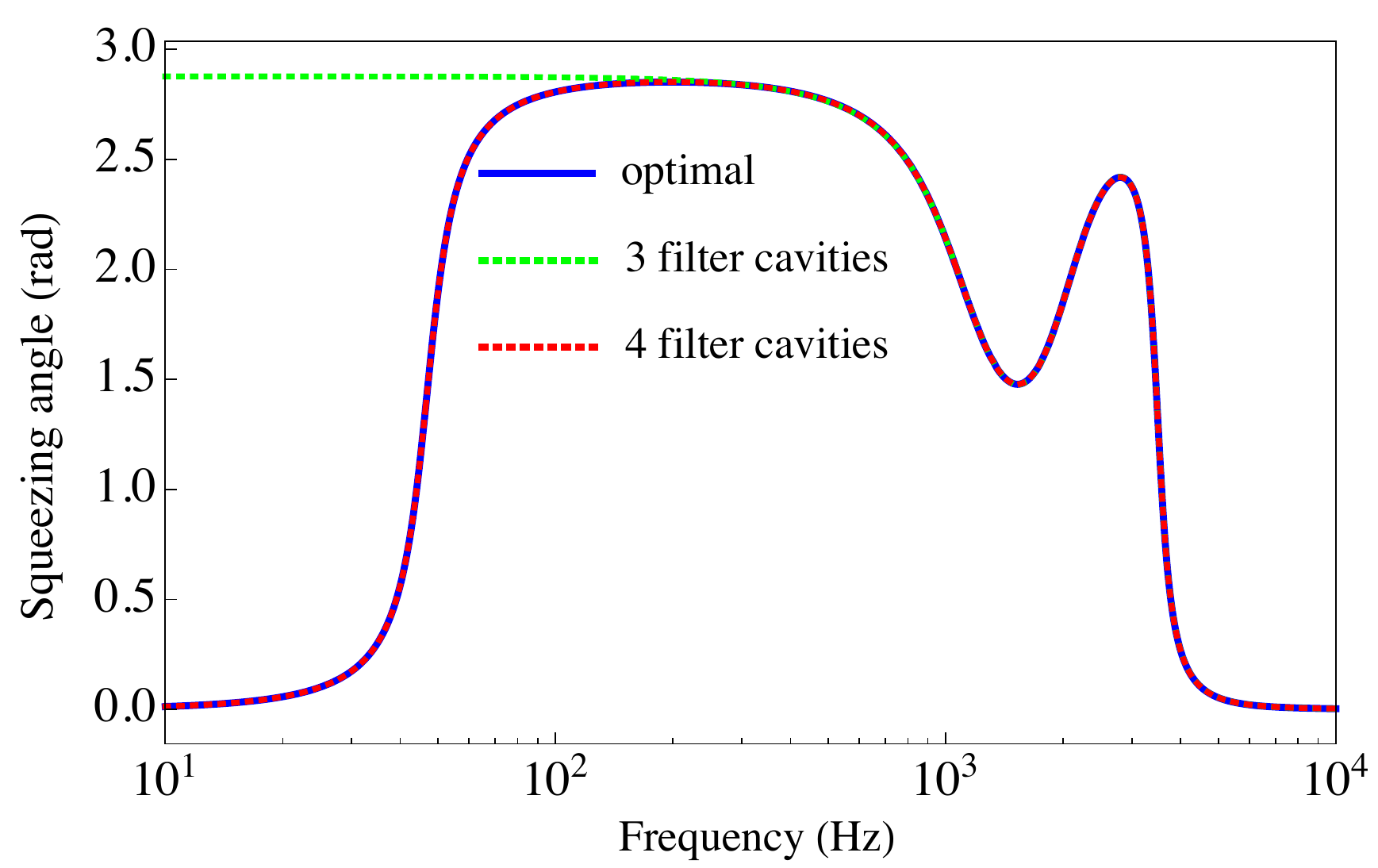}
\caption{The realization of frequency-dependent squeezing angle with 3 or 4 filter cavities. 
The 4-cavity realization matches the optimal one at all frequencies of interest, while the 
3-cavity realization matches the optimal one above 100 Hz. }
\label{fig:sqz}
\end{figure}

The transfer matrix ${\bf M}_{\rm OPT}$ of the optomechanical filter is the same as a detuned optomechanical
cavity, which has been extensively studied in the literature (see, e.g. the review 
article\,\cite{Chen2013, Aspelmeyer2014}). However, one usually only looks at a pair of 
sidebands around the laser frequency (equal to 
$\omega_0+\omega_m-\Delta_{\rm SR}$ in our case). 
Here we need to include two pairs: 1\&4, and 2\&3, which are 
pairs between signal and idler sidebands. When the filter cavity 
bandwidth is much smaller than the mechanical frequency, at the so-called 
resolved-sideband limit, the idler sidebands can be ignored, which can 
give rise to a rather simple input-output relation (like a negative 
bandwidth cavity) as shown in 
Ref.\,\cite{Miao2015a}. 
However, given the parameters that we have chosen, 
we are not in the ideal resolved-sideband limit. Interestingly, their influence 
on the signal sidebands can be
coherently suppressed as long as they are 
not resonant inside the interferometer, i.e., the accumulated propagation 
phase of $\omega_0+2\omega_m-2\Delta_{\rm SR}$ 
differs from integer multiples of $2\pi$, 
which is assumed in our analysis.

\begin{figure}[t]
\includegraphics[width=\columnwidth]{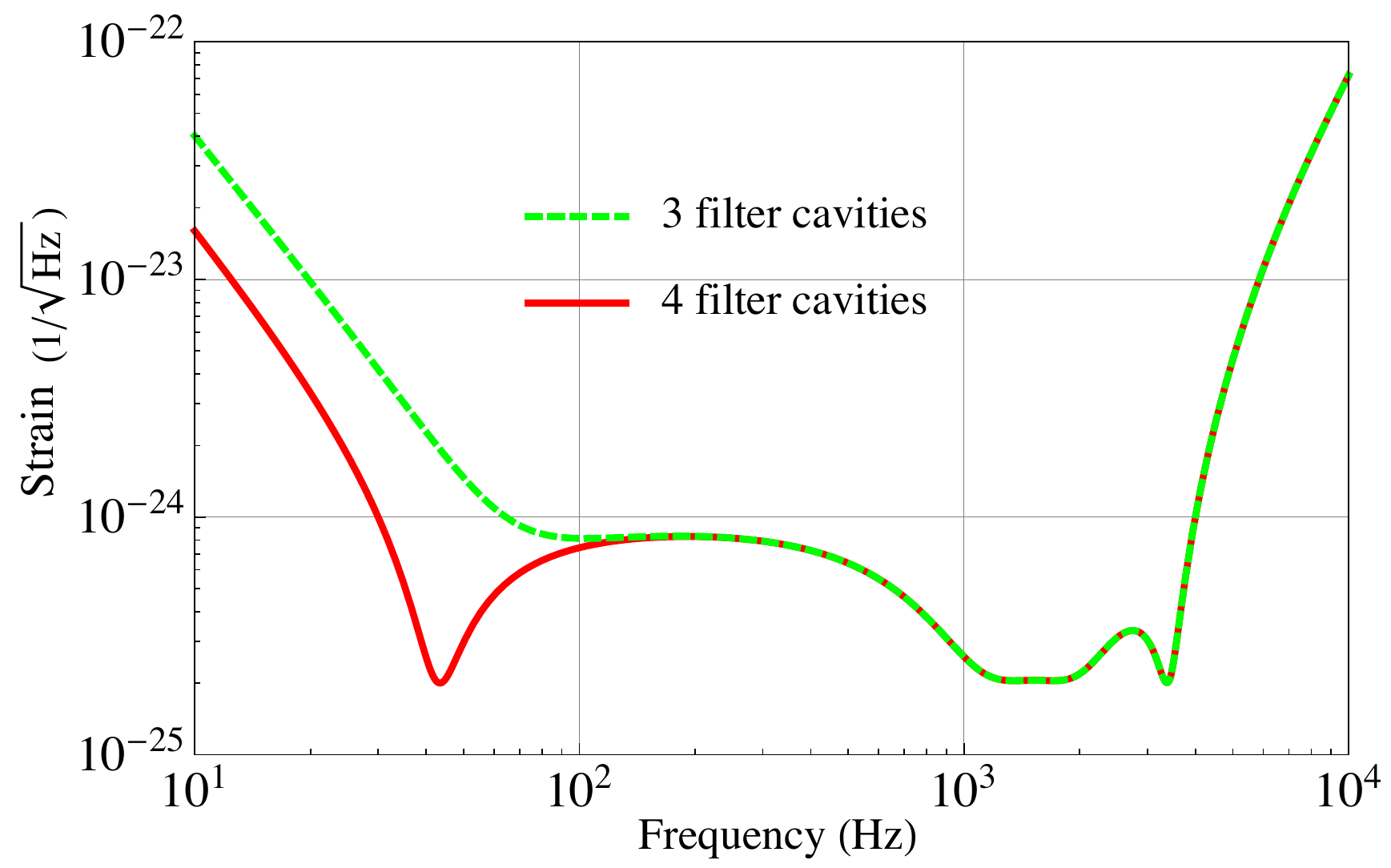}
\caption{The corresponding sensitivity curves (quantum noise only) for the two realisations with
3 and 4 squeezing filter cavities.}
\label{fig:sqz_sens}
\end{figure}

\begin{figure}[b]
\includegraphics[width=\columnwidth]{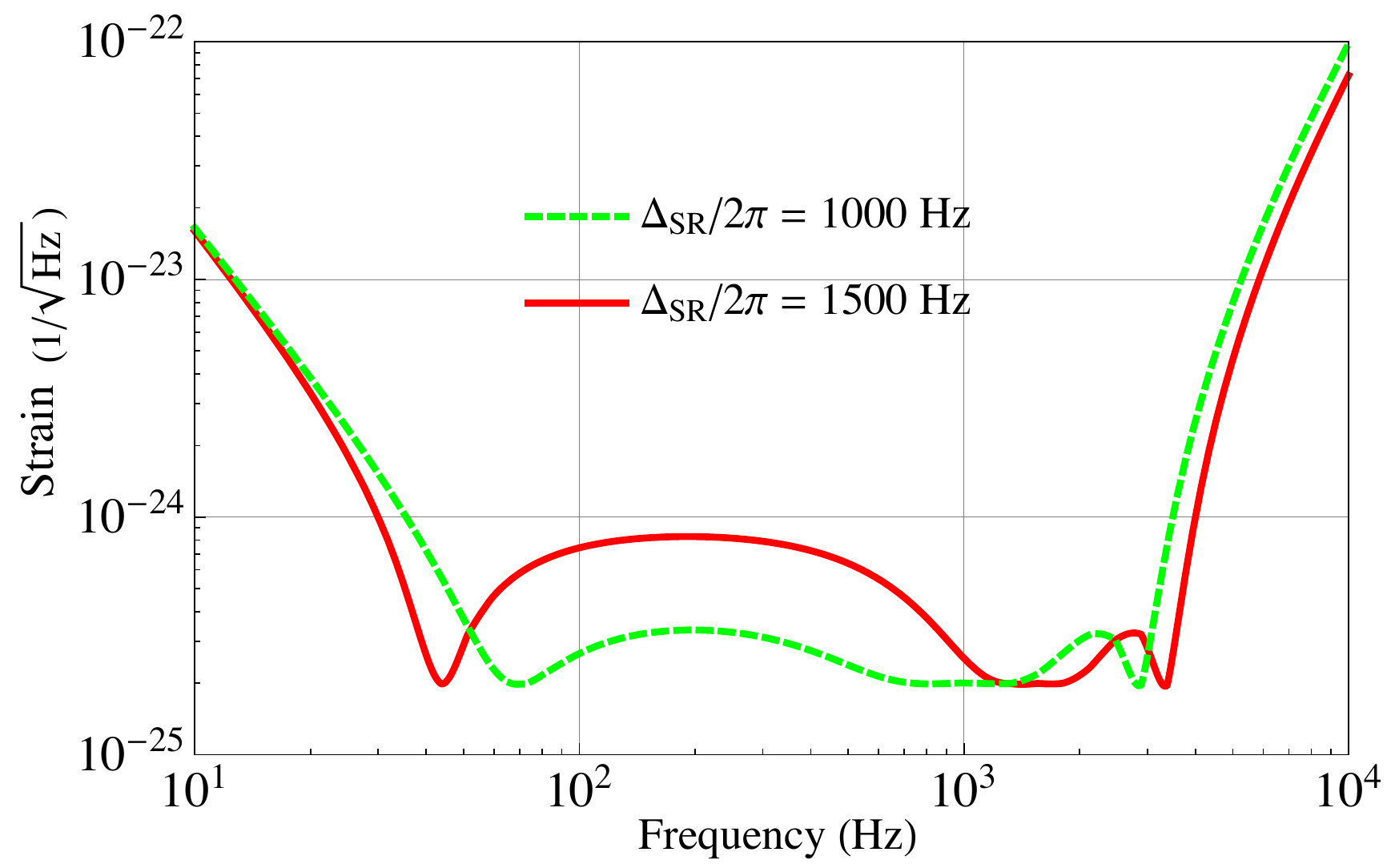}
\caption{Plot illustrating the effect of the signal-recycling detuning frequency $\Delta_{\rm SR}$ on the quantum-limited sensitivity. 
}
\label{fig:detune}
\end{figure}

\begin{figure}[t]
\includegraphics[width=\columnwidth]{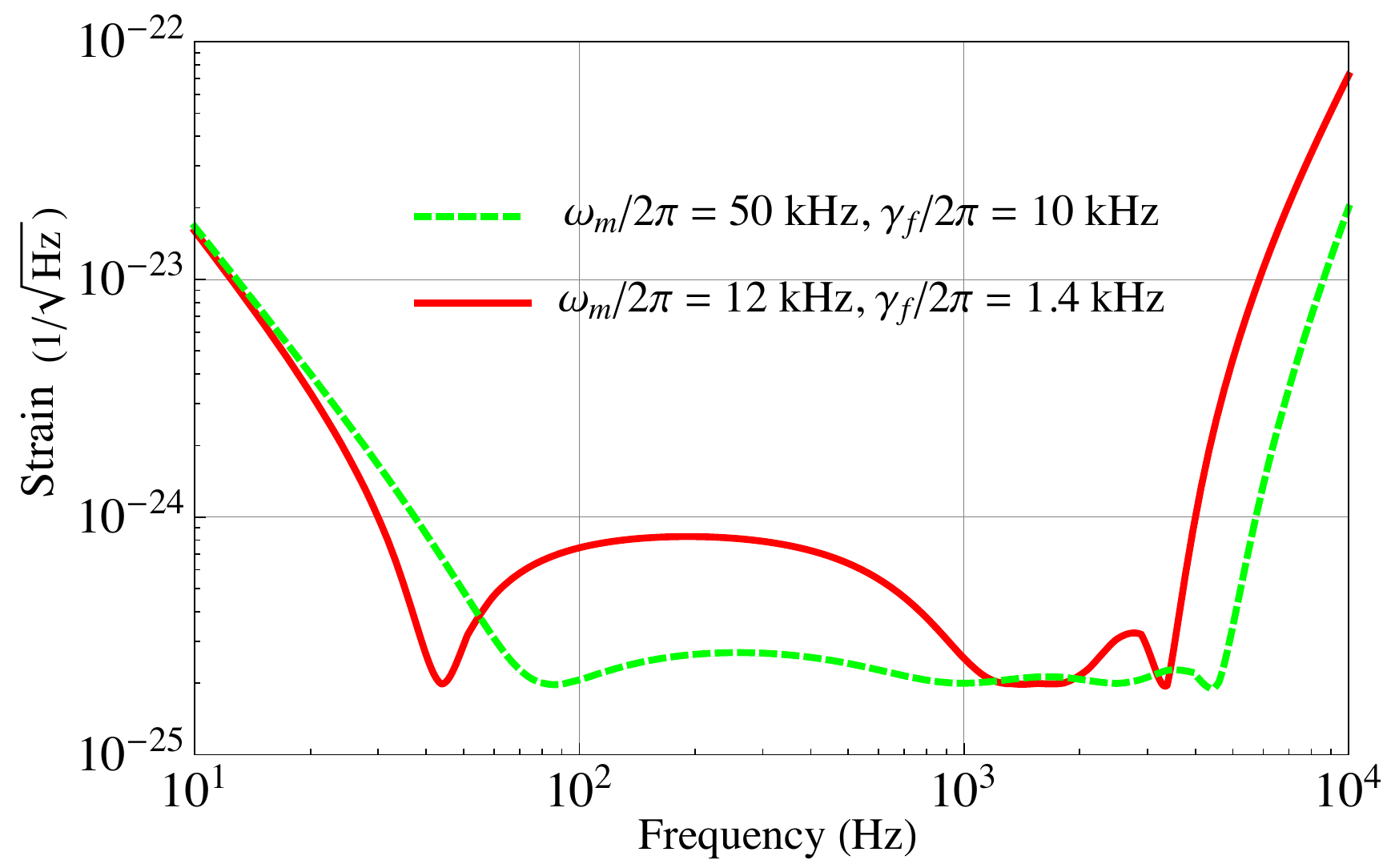}
\caption{Plot showing how different values of the mechanical resonant frequency $\omega_m$ influence the sensitivity. }
\label{fig:oscillator}
\end{figure}

\begin{figure}[b]
\includegraphics[width=\columnwidth]{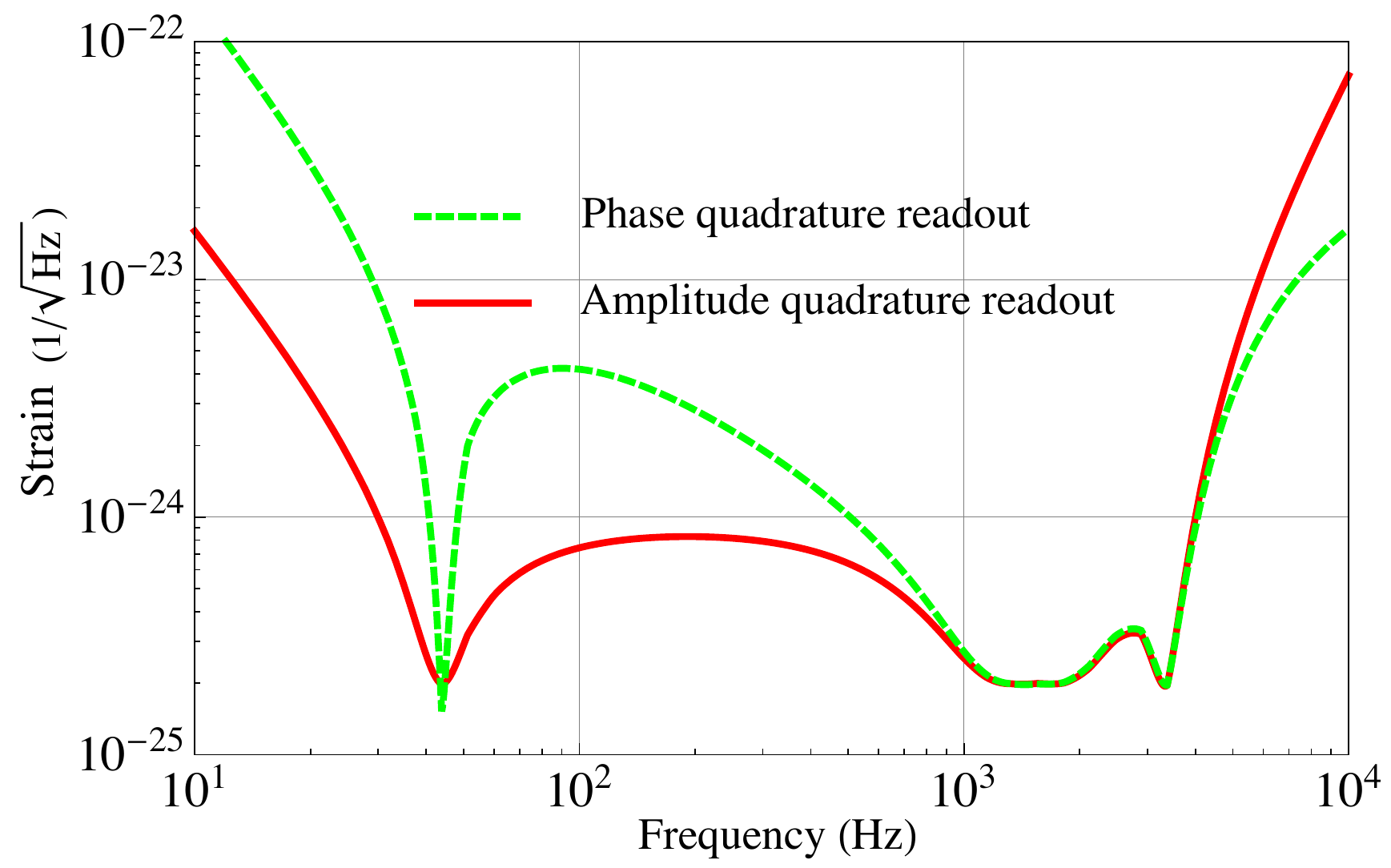}
\caption{Plot showing how choices of the readout quadrature affect the sensitivity. }
\label{fig:readout}
\end{figure}

The transfer matrix ${\bf M}_{\rm SQZ}$ for the squeezing has two parameters:
the squeezing factor and angle. We only include squeezing for the signal sidebands
and leave the idler sidebands in the vacuum state. The squeezing factor is assumed
to be 10dB observed at the final output; the squeezing angle is made to be
frequency dependent 
by sending the squeezed light through a cascade of filter cavities. The optimal 
frequency-dependent angle is derived by using the approach discussed 
in Ref.\,\cite{Harms2003}. The parameters for the filter cavities shown in 
Table I of the main text are obtained 
using a standard numerical fitting algorithm. 
As shown in Fig.\,\ref{fig:sqz}, we can match 
the optimal frequency-dependent angle with four filter cavities. 
If we focus on frequencies above 100 Hz, three filter cavities will be sufficient. 
Fig.\,\ref{fig:sqz_sens} shows the resulting sensitivity curves for these two cases. 

In Fig.\,\ref{fig:detune}, we show how the quantum-limited sensitivity changes with respect to
the signal-recycling detuning frequency. Pushing the high-frequency sensitivity by increasing the 
detuning frequency comes at the price of sacrificing the sensitivity at intermediate frequencies. We 
choose $\Delta_{\rm SR}/(2\pi) = 1.5$ kHz to get a good sensitivity up to 4\,kHz, and at
the same time, having the intermediate-frequency sensitivity close to the classical noise 
budget. 

In Fig.\,\ref{fig:oscillator}, we show the effect of mechanical oscillator frequency on the 
sensitivity. A higher oscillator 
frequency allows a large bandwidth $\gamma_f$ for the optomechanical filter
 while still remaining 
approximately within the resolved-sideband limit. 
Since $\gamma_f$ sets the upper limit of detector bandwidth, we can push the sensitivity 
curve to a higher frequency with  
a high-frequency oscillator. However, it is much more challenging to 
achieve 50 kHz using the
optical spring effect based upon the current technology. The 
power in the auxiliary cavity is nearly 20 times higher than the 12 kHz realization. 
That is why we have chosen 12 kHz as a compromise 
between the sensitivity and experimental feasibility. If there are
 other approaches to realizing a
low-loss mechanical oscillator at low temperatures, 
we could then consider high-frequency oscillators to achieve a better sensitivity. 

In Fig.\,\ref{fig:readout}, we show how the readout quadrature affects the sensitivity. 
In obtaining the sensitivity in Fig. 1 of the main text, we have assumed
 the measurement of amplitude quadrature to get a better
low-frequency sensitivity than the phase-quadrature readout. This 
 implies that we need a balanced homodyne detection scheme, which is planned to 
 be implemented in the near-term upgrade of the advanced detectors.

\section{More details on science cases}

\subsection{Neutron star binaries}

\begin{figure}[b]
\includegraphics[width=\columnwidth]{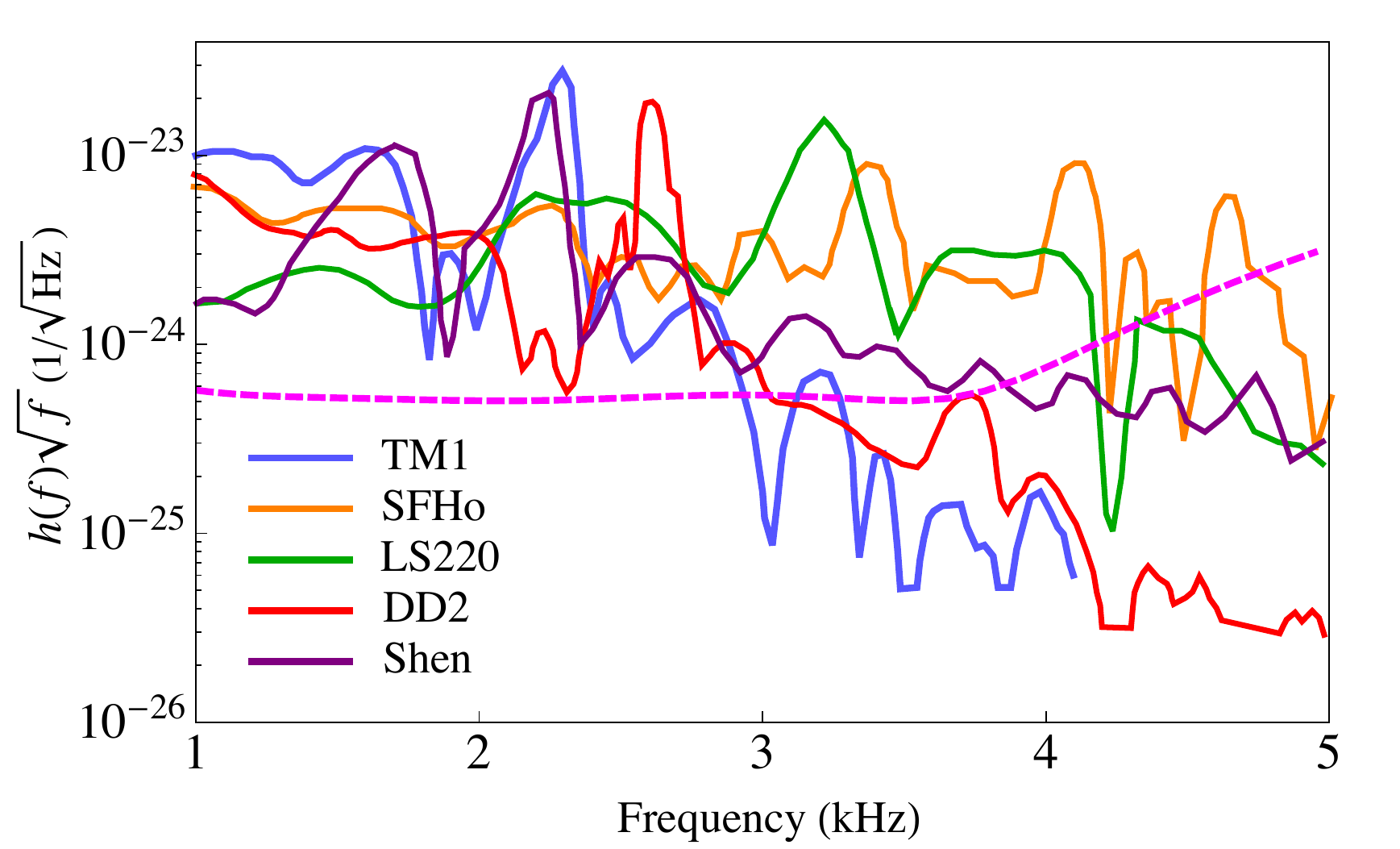}
\caption{BNS post-merger waveforms for different EOS extracted from Refs.\,\cite{Palenzuela:2015dqa,kiuchi,Stergioulas:2011gd}, assuming a $1.35 M_\odot+1.35 M_\odot$ NS binary located at $100\,{\rm Mpc}$ away. The magenta dashed line represents the target sensitivity of the high-frequency detector design. }
\label{fig:nwaveplot}
\end{figure}

In the main text, we have picked five representative EOSs that cover a range of 
stiffness, along with their corresponding maximum neutron star masses above $2M_\odot$. These  EOSs all take into account the finite-temperature effect self-consistently. For equal mass $1.35 M_\odot + 1.35 M_\odot$ neutron star binaries at a distance of $100\,{\rm Mpc}$, Fig.~\ref{fig:nwaveplot} shows the post-merger waveforms for different EOSs. Current numerical waveforms still contain significant theoretical uncertainties, and the physical origins of different oscillation modes are not completely understood. The dominant peak mode, associated with $\ell=2,m=2$ GW emission by the rotation of the remnant, is the most robustly determined in simulations. As a result, we pick this (2,2) mode as an example in the main text to illustrate the basic idea behind the post-merger spectroscopy. 

The waveform for the (2,2) mode can be approximated as a decaying sinusoid\,\cite{yang2017gravitational,bauswein2015exploring}:
\begin{align}\label{eq:do}
h(t) = A'\left( \frac{50 {\rm Mpc}}{d}\right)\sin (2 \pi f_{\rm peak} t-\phi_0) e^{-\pi f_{\rm peak} t/Q} \Theta(t)\,.
\end{align}
Here $d$ is the source distance, $A'$ is the amplitude, $f_{\rm peak}$ is the oscillation frequency, $Q$ is the quality factor of the oscillation, $\Theta(t)$ is the Heaviside function, and $\phi_0$ is the initial phase. We obtain the values for these parameters by fitting it to the numerical waveform in the frequency domain. The detailed procedure is explained in Ref.\,\cite{yang2017gravitational}. The fitted 
parameter values are listed in Table\,\ref{table:para}, with additional information on the star radius and threshold mass for a prompt collapse.

\begin{table}[t]
\centering
\begin{tabular}{c c c c c c}
\hline\hline
EOS & $R_{1.6 M_\odot}$ & $f_{\rm peak} (\rm kHz) \frac{M_\odot}{m_1+m_2}$ & $\frac{A' (50 {\rm Mpc})}{10^{-22}}$ & Q & $\frac{M_{\rm thres}}{M_\odot}$ \\
\hline
SFHo & 11.54 &1.25 & 2.7 & 25.7 & 2.95 \\
LS220 & 11.87 & 1.19 & 4.3 & 25.7 & 3.05 \\
DD2 & 13.66 & 0.93 &  2.8 & 12.7 & 3.35 \\
Shen & 14.65 & 0.82 & 5.0 & 23.3 & 3.45 \\
TM1 & 14.36 & 0.85 & 2.5 & 34.2 & 3.1 \\
\hline
\end{tabular}
\caption{The fitted values for the parameters in the analytically-approximated (2, 2) mode waveform for different EOSs.}
\label{table:para}
\end{table}

\begin{figure}[b]
\includegraphics[width=8.4cm]{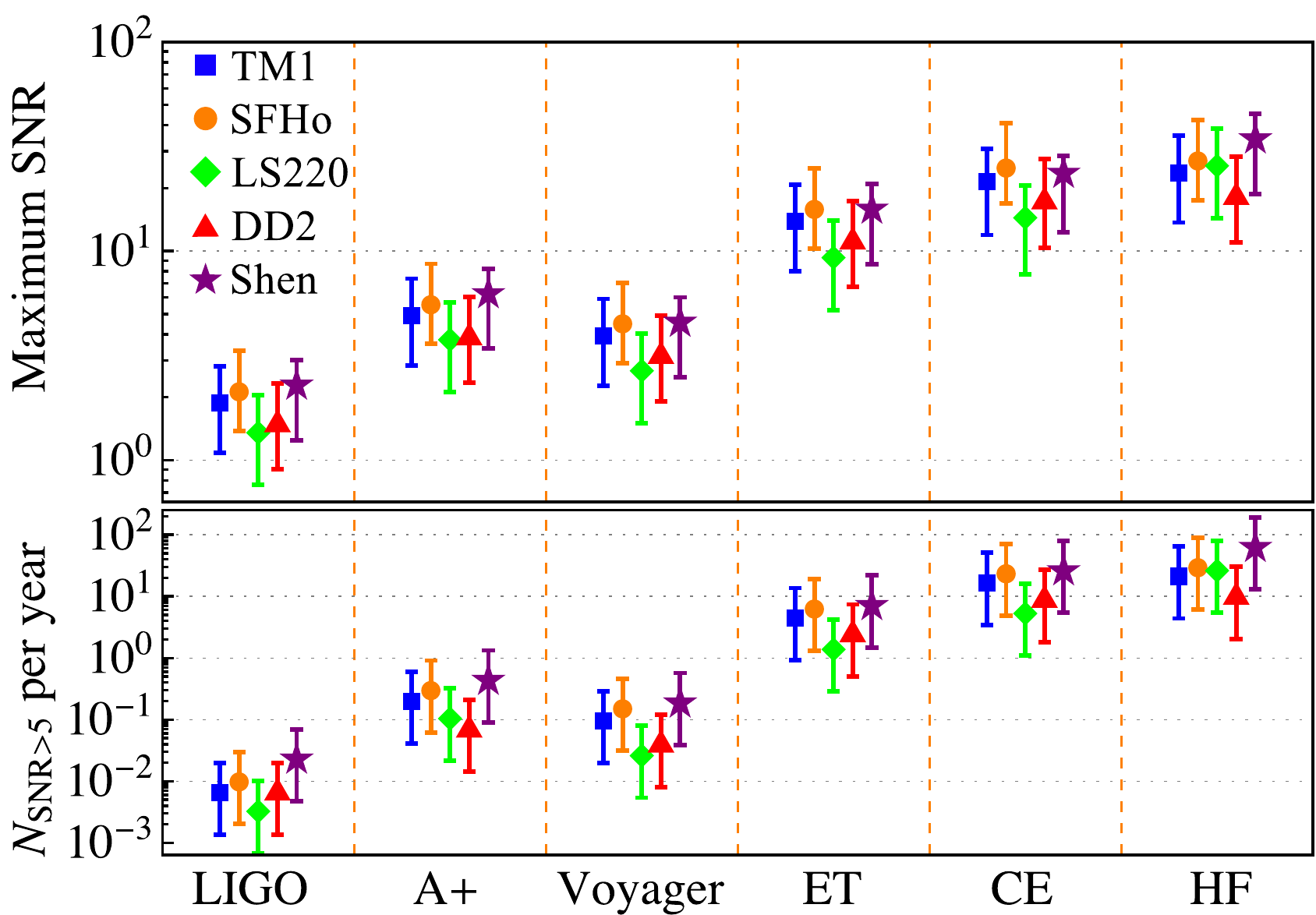}
\caption{Expected SNRs for detecting the entire post-merger waveform of the loudest event, i.e., the maximum SNR (top panel), and the number of events with ${\rm SNR} \ge 5$ (bottom panel) within a one-year observation for different detectors. Each EOS has a bar corresponding to the $90\%$ confidence interval of the merger rate obtained in Ref.\,\cite{GW170817}. The selected EOSs cover a range of stiffness with the corresponding NS mass above $2 M_\odot$.}
\label{fig:totalsnrplot}
\end{figure}

In performing the Monte-Carlo (MC) simulations, we have randomly sampled the distance (redshift $z \le 0.2$), sky location, polarization angle, and inclination angle of the BNS, which will affect the actual GW amplitude measured by a detector due to the antenna response\, \cite{sathyaprakash2009physics}. In addition, we assume that each component mass of the binary follows the distribution presented in Ref\,\cite{ozel2016masses}:
\begin{align}\label{eq:pm}
P(m_i; M_0,\sigma) =\frac{1}{\sqrt{2 \pi \sigma^2}} {\rm exp} \left [ -\frac{(m_i-M_0)^2}{2 \sigma^2}\right  ]\,,
\end{align}
with $M_0 =1.33 M_\odot$ and $\sigma =0.09 M_\odot$. We also adopt
the following fitting formula for the frequency of the (2, 2) 
mode\,\cite{bauswein2015exploring}:
\begin{align}\label{eq:fp}
\frac{f_{\rm peak}}{1  {\rm kHz}} 
= \left(\frac{m_1+m_2}{M_\odot} \right)\left [a_2 \left(\frac{R_{1.6 M_\odot}}{1 {\rm km}} \right)^2 +a_1 \left(\frac{R_{1.6 M_\odot}}{1 {\rm km}} \right)+a_0\right ]\,.
\end{align}
Here $a_0=5.503$, $a_1=-0.5495$ and $a_2=0.0157$ are EOS-independent parameters.
$R_{1.6 M_\odot}$ is the radius of a non-rotating NS with gravitational mass
$1.6M_\odot$, which encodes the EOS dependence.

For each EOS listed in Table~\ref{table:para}, we have performed $100$ MC realizations. For each, we assumed an one-year observation with the merger rate in Ref.\,\cite{GW170817}. 
We select the loudest events, and pick the median value of this set, which represents a $50\%$ percentile expectation of the maximum SNR. The SNRs for detecting the entire post-merger waveform, given the five EOSs, are presented in Fig.~\ref{fig:totalsnrplot}. 
These are different from Fig.\, 2 in the main text which shows the SNRs for detecting only the (2, 2) mode. The ratio between these two kinds of SNR is around a factor of 3 to 6, depending on the underlying EOS.

\subsection{Stochastic gravitational wave background}

The stochastic gravitational wave background (SGWB) is often characterized by:
\begin{align}\label{eqomega}
\Omega_{\rm GW}(f) \equiv \frac{1}{\rho_c} \frac{d \rho_{\rm GW}}{d \log f}=\frac{8\pi G}{3 c^2 H^2_0} \frac{d \rho_{\rm GW}}{d \log f}\,,
\end{align}
with $\rho_c$ being the critical energy density of the universe and $H_0 \approx 67.8 \,{\rm km\, s^{-1}} {\rm Mpc}^{-1}$ being the Hubble's constant.
The SGWB can be computed from the following formula: 
\begin{align}
\Omega_{\rm GW} = \frac{f}{\rho_c H_0} \int^{\infty}_0 dz\, \frac{R_m(z;\bm \theta) \frac{d E_{\rm GW}}{d f}(f_s ; \bm \theta)}{(1+z)\sqrt{\Omega_M(1+z)^3+\Omega_\Lambda}}\,,
\end{align}
where $\bm \theta$ are source parameters, $f_s=f(1+z)$ is the frequency at the source frame, $dE_{\rm GW}/df$ is the energy spectrum emitted by a single binary, $\Omega_M=1-\Omega_\Lambda =0.308$, and $R_m(z, \bm \theta)$ is the merger rate. In our analysis, we have assumed the nominal merger rate to be $1540 \,{\rm Gpc}^{-3} {\rm yr}^{-1}$; a higher (lower) merger rate will increase (decrease) $\Omega_{\rm GW}$ proportionally. 
The relation between $d E_{\rm GW}/df$ and GW amplitude $h$ is:
\begin{align}
\frac{d E_{\rm GW}}{d f} =\frac{4 \pi^2}{5}  f^2 h^2_{+,m}(f) d^2 =\frac{4 \pi^2}{5}  f^2 h^2_{\times,m}(f) d^2 \,,
\end{align}
where  $h_{+(\times), m}$ is the amplitude along the direction of maximal emission for 
a given polarization. 

There is an important subtlety to be highlighted. In principle, the source parameters $\bm \theta$ include the component masses, spins, etc. The total
$\Omega_{\rm GW}$ has to average over the distribution of them.
However, there is a limited number of numerical simulations for different EOSs, component masses, and spins. As a result, we have to make assumptions to proceed with the analysis. For producing Fig.\,3 in the main text, we have considered the simplest scenario and assumed the post-merger waveform of a $1.35 M_{\odot}-1.35 M_{\odot}$ binary for BNSs. 

\begin{figure}[t]
\includegraphics[width=\columnwidth]{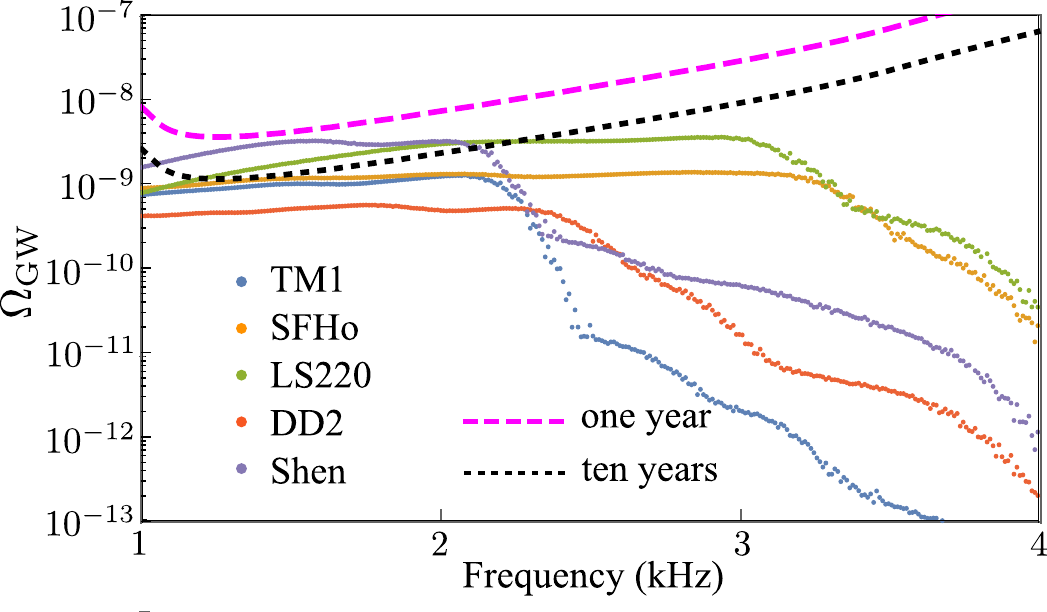}
\caption{SGWB from post-merger oscillations of BNS. 
The component masses are sampled according to
Eq.~\eqref{eq:pm} and the mode frequencies are scaled according to Eq.~\eqref{eq:s}. The scattered points come from the finite size of the Monte-Carlo sample.}
\label{fig:gwbplot2}
\end{figure}

Here, we investigate a slightly more complicated scenario by considering the component masses of the BNS following the distribution in Eq.~\eqref{eq:pm}. To model the mass dependence of the post-merger waveform, we assume that
 the amplitude of different modes in the post-merger stage for a generic binary with a total mass $m_1+m_2$ is the same as the ``canonical" $1.35M_\odot-1.35 M_\odot$ binary, and thus their frequencies are shifted according to: 
 \begin{align}\label{eq:s}
 f'_{\rm mode, i} = f_{\rm mode, i} \frac{m_1+m_2}{2.7 M_\odot}\,, 
 \end{align} 
which is a good approximation particularly for the (2, 2) mode, as shown in Eq.~\eqref{eq:fp}. 
However, it is not clear whether such scaling relation holds for the entire post-merger spectrum. 
Nevertheless, the resulting SGWBs for different EOSs are shown in Fig.~\ref{fig:gwbplot2}, assuming a one-year or ten-year 
observation with two co-located high-frequency detectors.
Comparing to those in Fig.\,3 in the main text, they are slightly smoother, which is due to the average over different
component masses, but the overall amplitude and qualitative behavior are very similar. This indicates that the SGWBs are robust against the variation of the component masses. In the future, the characterization of the SGWB can be improved with a better understanding of the post-merger waveform.

\subsection{Neutron star-black hole binaries}

The merger and ringdown waveforms of NS-BH mergers can be classified into three types\,\cite{shibata2009gravitational}, depending on the mass ratio of the two compact objects and the compactness of the NS. For the first type, the NS is disrupted outside the Innermost Stable Circular Orbit (ISCO), and the subsequent mass inflow produces weak GWs which ``shut off" quickly. The waveform can be well approximated by:
\begin{align}
h(f) =h_{\rm 3PN}(f) e^{-(f/f_{\rm cut})^\sigma}\,,
\end{align}
where $f_{\rm cut}$ and $\sigma$ are fitting parameters to the numerical waveforms.
For the second type, the NS is disrupted within the ISCO. The post-merger waveform, however, differs significantly from the ringdown waveform of binary BH mergers. For the third type, the NS is swallowed by the BH, and the post-merger waveform is close to the ringdown waveform of binary BH mergers. Type-II and Type-III waveforms can be parametrized by:
\begin{align}
h(f) =h_{\rm 3PN}(f) e^{-(f/f_{\rm ins})^\sigma}+\frac{A M}{d\, f} e^{-(f/f_{\rm cut})^{\sigma_{\rm cut}}}(1-e^{-(f/f_{\rm ins2})^5})\,,
\end{align}
where the parameters $f_{\rm ins}, f_{\rm ins2}, f_{\rm cut}, A, \sigma_{\rm cut}$ are fitting parameters, and $d$ is the source distance. 
 Following the discussions in Ref.\,\cite{shibata2009gravitational}, we assume that the post-merger process starts at $f_{\rm cut}$, which is the frequency at which the tidal disruption begins for Type-I and Type-II waveforms. For type-III waveform, tidal disruption never happens, and $f_{\rm cut}$ is close to the quasinormal frequency of the final black hole.

\begin{table}[t]
\centering
\begin{tabular}{c c c c c c c}
\hline\hline
Type & $M_{\rm BH}/M_{\rm NS}$ & $f_{\rm ins} M$ & $f_{\rm ins2} M$ & $A$ & $f_{\rm cut} M$ & $\sigma_{\rm cut}$ \\
\hline
I &1.5 &  & &  & 0.03 & 2.2 \\
I & 2 &  &  &  & 0.038 & 2.2 \\
II & 3 & 0.014 & 0.014 &  0.13 & 0.063 & 2.9 \\
III & 4 & 0.016 & 0.016 & 0.103 & 0.079 & 4.6 \\
III & 5 & 0.019 & 0.02 & 0.09 & 0.087 & 3.7 \\
\hline
\end{tabular}
\caption{Parameters for different types of BH-NS mergers}
\label{table:bhns}
\end{table}

In Table~\ref{table:bhns} we list the values of BH-NS waveform parameters obtained from Ref.\,\cite{shibata2009gravitational} and 
used for generating Table II in the main text.  
Here, $M$ is the total mass of the binary: $M=M_{\rm BH}+M_{\rm NS}$, with the NS mass fixed to be $1.35 M_\odot$
and the compactness fixed to be $M_{\rm NS}/R_{\rm NS} =0.145$, assuming a $\Gamma=2$ polytropic EOS.

\subsection{Measuring $H_0$ without electromagnetic counterparts}

BNS mergers can be used to infer the Hubble constant without electromagnetic counterparts, either by statistically identifying the host galaxies with a catalogue of events\,\cite{schutz1986determining}, or performing mass distribution reconstruction\,\cite{PhysRevD.85.023535}, or through the tidal-effect measurement in the inspiral stage assuming that the EOS is known\,\cite{messenger2012measuring}. Here, we follow the discussion in Ref.\,\cite{Messenger:2013fya} to illustrate an alternative approach to obtaining the redshift information by comparing the inspiral and post-merger waveforms. 

Since only close BNS events will have their post-merger waveforms detected, we can use $z=H_0 d$ to estimate the accuracy of measuring $H_0$, namely, 
\begin{align}
\delta H_0 = H_0 \sqrt{\left (\frac{\delta z}{z}\right )^2+\left (\frac{\delta d}{d}\right )^2} \equiv \sqrt{(\delta H^z_0)^2+(\delta H^d_0)^2}\,.
\end{align}
For these events, the accuracy in the distance measurement should reach a percent level of precision, which is much smaller than the redshift-related uncertainty $\delta H^z_0$.
We, therefore, can approximate $\delta H_0$ as $\delta H^z_0$.

The post-merger waveforms for BNS with the same mass ratio but different total mass are different. The redshift scaling on mass in the inspiral stage no longer holds here, and this is the key to break the degeneracy in redshift. Because there are significant uncertainties in modeling the entire post-merger waveform, we only discuss how to use the dominant peak (2, 2) mode to infer the redshift. In the future, when the modelling is improved, the entire post-merger waveform can be used to obtain a more precise estimation of $H_0$.

For the (2, 2) mode, if the EOS is known, its frequency can be inverted to obtain the total mass using Eq.~\eqref{eq:fp}:
\begin{align}
\frac{M}{M_\odot} =  \frac{(1+z) f_{\rm peak,m}}{1 {\rm kHz}}\left [a_2 \left(\frac{R_{1.6 M_\odot}}{1 {\rm km}} \right)^2 +a_1\left( \frac{R_{1.6 M_\odot}}{1 {\rm km}} \right)+a_0\right ]^{-1}\,, 
\end{align}
in which $f_{\rm peak,m}$ is the measured redshifted frequency on Earth. 
Additionally, the redshifted chirp mass $\mathcal{M}_z$ can be estimated using the inspiral waveform with a high precision:
\begin{align}
\mathcal{M}_z =(1+z)M \eta^{5/3}\,,
\end{align}
where $\eta =m_1 m_2/M^2$ is the symmetric mass ratio, and we approximate it as a known number here.

By comparing the above two equations, we find that:
\begin{align}
\frac{\delta f_{\rm peak,m}}{f_{\rm peak,m}}\approx \frac{\delta (1+z)^2}{(1+z)^2}  \approx  2 \delta z\,.
\end{align}
In other words,
\begin{align}\label{eqdzz}
\frac{\delta z}{z} =\frac{\delta H^z_0}{H_0} & \approx \frac{\delta f_{\rm peak,m}}{f_{\rm peak,m}} \frac{1}{2 H_0 d }\,\nonumber \\
& \approx \frac{0.7}{Q\, {\rm SNR}} \frac{1}{2 H_0 d }\,.
\end{align}
where the last line uses the result in Ref.\,\cite{yang2017gravitational}, $Q$ is the quality factor of the (2, 2) mode, which depends on the EOS. Since the ${\rm SNR}$ for measuring the (2, 2) mode is proportional to $1/d$, this implies that $\delta z/z$ is independent of the source distance. 

We can evaluate Eq.~\eqref{eqdzz} for different EOSs (with the averaged SNR), and obtain:
\begin{equation}
\frac{\delta H^z_0}{H_0} \equiv N^{-1/2} \times \left\{
\begin{array}{cl}
10\%\,, & {\rm TM1}\,, \\
\\
17\%\,, & {\rm SFHo}\,,\\
\\
10\%\,, & {\rm LS220}\,,\\
\\
39\%\,, & {\rm DD2}\,,\\
\\
8\%\,, & {\rm Shen}\,,
\end{array}\right.
\label{bx}
\end{equation}
where $N$ is the number of events. The performance of this approach crucially depends on the quality factor and amplitude of the mode. For most EOSs listed here, the measurement accuracy on $H_0$ after one-year observation is comparable to, if not better than, those using other methods\,\cite{messenger2012measuring}.
To include low-SNR events, the above Fisher-type estimation has to be replaced by a proper Bayesian analysis in which the posterior distribution of $H_0$ can be reconstructed based on multiple events\,\cite{Messenger:2013fya}.

\subsection{Other astrophysical sources}

In addition to those sources discussed earlier, an interesting source is the rotating and oscillating NSs in our galaxy.
For example, rapidly rotating pulsars may generate GWs above $1 \,{\rm kHz}$. Moreover, isolated NSs have a family of modes in the high-frequency range\,\cite{Andersson1998}. The f-modes of an NS could be excited during a violent process, e.g., a giant flare event from a magnetar. Following the analysis in Ref.\,\cite{levin2011excitation},
the SNR of such an f-mode oscillation using the high-frequency GW detector can be estimated as:
\begin{align}
{\rm SNR} \approx & 1.2\, \Lambda \left ( \frac{2\, {\rm kHz}}{f_{\rm mode}}\right )^2 \left ( \frac{\sqrt{S_{hh}}}{5.0\times 10^{-25} {\rm Hz}^{-1/2}}\right )\left ( \frac{B}{10^{15} {\rm G}}\right )^2\nonumber \\
& \left ( \frac{1 {\rm kpc}}{d}\right ) \left ( \frac{R}{10 {\rm km}}\right )^2 \left ( \frac{0.07 M_\odot}{m_{\rm eff}}\right )^{1/2}\,, 
\end{align}
 where $f_{\rm mode}$ is the frequency of the f-mode, $m_{\rm eff}$ is the effective mass of the mode, $R$ is the star radius,
 $B$ is the strength of the external magnetic field, and $\Lambda$ is the overlapping function between the mode and its external driving, which is of the order of unity. Since the closest magnetar flare observed so far SGR $1900+14$ is at a distance
 of around $6.0 \, {\rm kpc}$, detecting the induced mode oscillation with $\rm SNR>1$ would either require a closer event or a further improvement of detector sensitivity, e.g., from increasing the arm length. Note that there are three magnetars observed in the past 40 years, so the time separation between giant flares for a given active magnetar can be estimated as around 40 years. 
This type of source is interesting because the GWs could be accompanied by energetic X-ray emissions\,\cite{thompson2017global}, which allows us to infer the details about the star's internal dynamics. 

Another important source is the core-collapse supernovae. Even though the dominant spectral power resides at frequencies below $1 \,\rm kHz$, the high-frequency modes also contain crucial information about the post-bounce dynamical process. By looking at Fig.\,8 in Ref.\,\cite{Ott:SN2009}, we can approximate the strain of high-frequency ($>1 \,{\rm kHz}$) GWs as
\begin{align}
h \approx 10^{-20} \,\beta \left ( \frac{10 \,{\rm kpc}}{d}\right ) \left ( \frac{f}{1\, {\rm kHz}} \right )^{-\alpha}\,,
\end{align}
where $\alpha$ is between $3$ and $4$, and $\beta$ is of the order of unity. The corresponding SNR for detecting the entire waveform above 1\,kHz is then approximately given by 
\begin{align}
{\rm SNR} \approx 6\, \beta \left(\frac{4}{\alpha+1}\right) \left ( \frac{1 \,{\rm Mpc}}{d}\right ) \left( \frac{\sqrt{S_{hh}}}{5.0\times 10^{-25} {\rm Hz}^{-1/2}}\right )\,.
\end{align}

\bibliography{references}

\end{document}